\documentclass[a4paper, oneside, reqno, 11pt]{amsart}
\usepackage{lipsum}
 \let\counterwithin\relax \usepackage{graphicx,amsfonts,amssymb,amsmath,amsthm,url,amscd} \usepackage{enumitem} \usepackage{color} \usepackage{chngcntr}
\usepackage[dvipsnames]{xcolor} \usepackage[normalem]{ulem}
\usepackage{ninecolors}
\usepackage{comment}
\usepackage{graphicx, amsmath, amssymb, amsfonts, amsthm, tikz, amscd,mathtools}
\usetikzlibrary{matrix,arrows,decorations.pathmorphing}
\usepackage[a4paper, margin=2.7cm]{geometry}
\usepackage{verbatim}
\usepackage{calc}
\usepackage{dsfont}

\usepackage[linesnumbered, ruled, vlined, algosection]{algorithm2e}
\SetKwInput{KwComplexity}{Complexity}
\SetKw{Continue}{continue}
\SetKw{False}{false}
\SetKw{True}{true}
\SetKw{And}{and}
\SetKw{Or}{or}
\SetKw{Delete}{delete}
\SetKw{Not}{not}
\SetKw{Then}{then}
\SetKw{PythFor}{ for }
\SetKw{St}{such that}
\SetKwRepeat{DoWhile}{do}{while}

\counterwithin{figure}{section}

\usepackage{placeins}

\let\Oldsection\section
\renewcommand{\section}{\FloatBarrier\Oldsection}

\let\Oldsubsection\subsection
\renewcommand{\subsection}{\FloatBarrier\Oldsubsection}

\usepackage{fewerfloatpages}

\usepackage{tabularray}

\usepackage[font = footnotesize]{caption}

\UseTblrLibrary{siunitx}
\UseTblrLibrary{counter}
\UseTblrLibrary{functional}

\DefTblrTemplate{caption-tag}{caption-ams-style}{\textsc{Table~\thetable .}}
\DefTblrTemplate{caption-sep}{caption-ams-style}{\enskip}
\DefTblrTemplate{caption-text}{caption-ams-style}{\InsertTblrText{caption}}
\DefTblrTemplate{conthead-text}{caption-ams-style}{(Continued~on~next~page)}

\ExplSyntaxOn
\DefTblrTemplate{caption}{caption-ams-style}{\vspace{0.7em}
	\hbox_set:Nn \l__tblr_caption_box
	{
		\UseTblrTemplate { caption-tag } { default }
		\UseTblrTemplate { caption-sep } { default }
		\UseTblrTemplate { caption-text } { default }
	}
	\dim_compare:nNnTF { \box_wd:N \l__tblr_caption_box } > { 0.85\linewidth }
	{
		\noindent
		\makebox[\linewidth][c]{
			\parbox{0.85\textwidth}{
				\hbox_unpack:N \l__tblr_caption_box
			}
		}
		\par
	}
	{
		\centering
		\makebox [\linewidth] [c] { \box_use:N \l__tblr_caption_box }
		\par
	}
}
\DefTblrTemplate{capcont}{caption-ams-style}{\vspace{0.7em}
	\hbox_set:Nn \l__tblr_caption_box
	{
		\UseTblrTemplate { caption-tag } { default }
		\UseTblrTemplate { caption-sep } { default }
		\UseTblrTemplate { caption-text } { default }
		\UseTblrTemplate { conthead-pre } { default }
		\UseTblrTemplate { conthead-text } { default }
	}
	\dim_compare:nNnTF { \box_wd:N \l__tblr_caption_box } > { 0.85\linewidth }
	{
		\noindent
		\makebox[\linewidth][c]{
			\parbox{0.85\textwidth}{
				\hbox_unpack:N \l__tblr_caption_box
			}
		}
		\par
	}
	{
		\centering
		\makebox [\linewidth] [c] { \box_use:N \l__tblr_caption_box }
		\par
	}
}
\ExplSyntaxOff

\DefTblrTemplate{firsthead,middlehead,lasthead}{caption-bottom-ams-style}{
}
\DefTblrTemplate{firstfoot}{caption-bottom-ams-style}{
	\UseTblrTemplate{capcont}{caption-ams-style}
}
\DefTblrTemplate{middlefoot}{caption-bottom-ams-style}{
	\UseTblrTemplate{capcont}{caption-ams-style}
}
\ExplSyntaxOn
\DefTblrTemplate{lastfoot}{caption-bottom-ams-style}{
	\UseTblrTemplate{note}{default}
	\UseTblrTemplate{remark}{default}
	\int_compare:nNnTF { \l__tblr_table_page_int } = {1} {
		\UseTblrTemplate{caption}{caption-ams-style}
	} {
		\UseTblrTemplate{capcont}{caption-ams-style}
	}
}
\ExplSyntaxOff

\NewTblrTheme{ams-theme}{
	\SetTblrTemplate{caption-tag}{caption-ams-style}
	\SetTblrTemplate{caption-sep}{caption-ams-style}
	\SetTblrTemplate{caption-text}{caption-ams-style}
	
	\SetTblrTemplate{conthead-text}{caption-ams-style}
	\SetTblrTemplate{caption}{caption-ams-style}
	
	\SetTblrTemplate{firsthead,middlehead,lasthead,firstfoot,middlefoot,lastfoot}{caption-bottom-ams-style}
}

\counterwithin{table}{section}

\usepackage{etoolbox} \patchcmd{\section}{\scshape}{\bfseries}{}{} \makeatletter \renewcommand{\@secnumfont}{\bfseries} \makeatother
\newrobustcmd\TableBold{\DeclareFontSeriesDefault[rm]{bf}{b}\bfseries}

\usepackage{graphicx, amsmath, amssymb, amsfonts, amsthm, stmaryrd, tikz, amscd} \usepackage[all]{xy} \usetikzlibrary{matrix,arrows,decorations.pathmorphing}

\usepackage[hyperfootnotes=false, colorlinks, citecolor= RoyalBlue, urlcolor=blue, linkcolor=blue ]{hyperref}

\makeatletter
\let\orgdescriptionlabel\descriptionlabel
\renewcommand*{\descriptionlabel}[1]{%
  \let\orglabel\label
  \let\label\@gobble
  \phantomsection
  \protected@edef\@currentlabel{#1}%
  \let\label\orglabel
  \orgdescriptionlabel{#1}%
}
\def\paragraph{
	\@startsection{paragraph}{4}
	\z@{.5\linespacing\@plus.7\linespacing}{-.5em}%
	{\normalfont\itshape}}
\makeatother
\theoremstyle{plain} %--default
\newtheorem{theorem} {Theorem}

\theoremstyle{definition}

\theoremstyle{plain} %--default
\newtheorem{proposition}[theorem] {Proposition}

\theoremstyle{remark}

\newtheorem*{definition*} {Definition}

 \newtheorem*{example*} {Example}

\newtheorem{remark}[theorem] {Remark} \newtheorem*{remark*} {Remark}       

\newtheoremstyle{itplain} % name
{6pt} % Space above
{5pt\topsep} % Space below
{\itshape} % Body font
{} % Indent amount
{\itshape} % Theorem head font
{.}  % Punctuation after theorem head
{5pt plus 1pt minus 1pt} % Space after theorem head
{} % Theorem head spec (can be left empty, meaning ‘normal’)

\theoremstyle{itplain} %--default
\newtheorem*{lemma*}{Lemma}

\newtheorem{corollary}[theorem]{Corollary}
\newtheorem*{corollary*} {Corollary}

\theoremstyle{remark} %--default

\newtheorem*{lemmatest*}{Lemma}

\usepackage{etoolbox} \patchcmd{\section}{\scshape}{\bfseries}{}{} \makeatletter \renewcommand{\@secnumfont}{\bfseries} \makeatother

\newlength{\faktorheight}

\makeatletter
\providecommand{\leftsquigarrow}{%
	\mathrel{\mathpalette\reflect@squig\relax}%
}
\newcommand{\reflect@squig}[2]{%
	\reflectbox{$\m@th#1\rightsquigarrow$}%
}
\makeatother

\numberwithin{theorem}{section}
\numberwithin{equation}{section}                                 \def\eps{\varepsilon}                        

                                     \DeclareMathOperator{\rank}{rank}

\DeclareMathOperator*{\argmin}{argmin}
\DeclareMathOperator{\select}{select}

\newcommand{\sm}{\left(\begin{smallmatrix}} \newcommand{\esm}{\end{smallmatrix}\right)} \newcommand{\bpm}{\begin{pmatrix}} \newcommand{\ebpm}{\end{pmatrix}}

\newcommand{\vect}[1]{{\boldsymbol{#1}}}

\newcommand{\lquotient}[2]{ \mathchoice {% \displaystyle
    \text{\lower1ex\hbox{$#1$}\Big \backslash \raise01ex\hbox{$#2$}}%
  } {% \textstyle
    #1\,\backslash\,#2 } {% \scriptstyle
    #1\,\backslash\,#2 } {% \scriptscriptstyle
    #1\,\backslash\,#2 } }

\newcommand{\rquotient}[2]{ \mathchoice {% \displaystyle
    \text{\raise01ex\hbox{$#1$}\Big/\lower1ex\hbox{$#2$}}%
  } {% \textstyle
    #1\,/\,#2 } {% \scriptstyle
    #1\,/\,#2 } {% \scriptscriptstyle
    #1\,/\,#2 } }
		
\newcommand{\lrquotient}[3]{ \mathchoice {% \displaystyle
    \text{\lower1ex\hbox{$#1$}\Big \backslash \raise01ex\hbox{$#2$}\Big/\lower1ex\hbox{$#3$}}%
  } {% \textstyle
    #1\,\backslash\,#2\,/\,#3 } {% \scriptstyle
    #1\,\backslash\,#2\,/\,#3 } {% \scriptscriptstyle
    #1\,\backslash\,#2\,/\,#3 } }

\author{Raphael S. Steiner}
\address{Huawei Research Center Zurich, Computing Systems Lab, Thurgauerstrasse 80, 8050 Z{\"u}rich, Switzerland} \email{raphael.steiner@huawei.com}

\author{Mirko De Vita}
\address{Huawei Research Center Zurich, Computing Systems Lab, Thurgauerstrasse 80, 8050 Z{\"u}rich, Switzerland} \email{mirko.de.vita@h-partners.com}

\author{Endri Bezati}
\address{Huawei Research Center Zurich, Computing Systems Lab, Thurgauerstrasse 80, 8050 Z{\"u}rich, Switzerland} \email{endri.bezati@huawei.com}

\date{\today} \title{Optimal tables for asymmetric numeral systems}
\hypersetup{ pdfkeywords={}, pdfsubject={}, pdfcreator={TeXstudio 2.12.22}}
\begin{document}

\begin{abstract} We present several algorithms to generate tables for asymmetric numeral systems and prove that they are optimal in terms of discrepancy. In turn, this gives rise to the strongest proven bound on entropy loss. We further give improved theoretical bounds for the entropy loss in tabled asymmetric numeral systems and a brief empirical evaluation of the stream variant.
\end{abstract}

\maketitle

\setcounter{tocdepth}{1} \tableofcontents

\maketitle

\section{Introduction}\label{sec:introduction}

Asymmetric numeral systems were introduced by Duda \cite{DudaANS09,DudaANS13,DudaTahboubGadgilDelp15} as a coding algorithm. It is a lossless compression algorithm that offers the speed of Huffman coding \cite{HuffmanCoding} and the compression rates of arithmetic coding \cite{rissanen1976generalized,langdon1982simple,langdon1984introduction,witten1987arithmetic}. As such, it has drawn wide-ranging attention from research \cite{GiesenInterleaved, yokoo2018probability, DudaANS21, duda2023lightweight, pieprzyk2023compression} and has been adopted by industry: Zstandard by Facebook \cite{Zstandard}, LZFSE by Apple \cite{LZFSE}, and PIK by Google \cite{pik}.

Besides compression, asymmetric numeral systems display chaotic behaviour: the change of a single bit in the encoded data yields completely different outputs when decoded. It has thus been put forth as a means to prove ownership of digital media such as non-fungible tokens (NFTs) \cite{hsieh2022review} and as a means of lightweight encryption \cite{camtepe2021compcrypt, mahboubi2022digital, cryptoeprint:2022/005, duda2023lightweight}.

For a non-empty finite set of symbols $\mathcal{S}$ and an everywhere positive probability measure $f$ on $\mathcal{S}$, asymmetric-numeral-system encoding aims to add a new symbol $s \in \mathcal{S}$ to an already encoded message $n \in \mathbb{Z}_{\ge 0}$ by replacing the encoded message with $n' \in \mathbb{Z}_{\ge 0}$, where $n' \approx \frac{n}{f(s)}$. This guarantees that the added information content of a single symbol is close to the optimal Shannon entropy \cite{shannon1948mathematical}. For example, if all symbols are equally likely, that is $f \equiv {1} \slash {|\mathcal{S}|}$, one might take the $|\mathcal{S}|$-ary expansion of $n$ and tuck the number in $\{ 0,1,\dots, |\mathcal{S}|-1 \}$ corresponding to the symbol $s$ at the end.

In general, asymmetric numeral systems use an allocation $A: \mathbb{Z}_{\ge 0} \to \mathcal{S}$ of numbers to symbols to generate the new encoded message. More precisely, the new encoded message $C(s, n) \in \mathbb{Z}_{\ge 0}$ is the index of the $(n+1)$-th occurrence of the symbol $s$ in the allocation $A$. Thus, the allocation driving the encoding is ultimately responsible for the efficacy of the compression in terms of entropy. Several strong algorithms were already introduced and analysed in Duda's initial paper \cite{DudaANS09}. First, for computational efficiency, the allocations were restricted to a repeating pattern such that the information could be easily stored in a small table. Furthermore, two different heuristics were given to generate the table:
\begin{enumerate}[label=(\roman*)]
	\item group the same symbols into a contiguous interval for computational efficiency; \label{item:duda-first} %see also \S\ref{sec:range-table}, 
	\item put the symbols in the same order as the symboled points 
	\begin{equation}
	\left\{\left(\tfrac{n}{f(s)},s\right) \in \mathbb{Q} \times \mathcal{S} \, | \, n \in \mathbb{Z}_{>0}, s \in \mathcal{S} \right\}
	\end{equation}
	appear on the number line. \label{item:duda-second} % see also \S\ref{sec:Duda-table}.
\end{enumerate}

We note that in order for there to be no bias, the densities of the symbols $s \in \mathcal{S}$ in the table need to match their probabilities $f(s)$, which we henceforth assume. If the probabilities cannot be match up with the densities, additional considerations have to be made, such as which symbols should be over- and under-represented in the table, and the placement should take both the probability and the density into account. We refer to \cite{DudaANS21} and references therein for such considerations in this case.

For an allocation of table length $Q$ generated by \ref{item:duda-first}, one easily proves, see Proposition \ref{prop:disc-trivial-bound}, that the index $N_s-1$ of the $n$-th occurrence of a symbol $s$ satisfies
\begin{equation} \label{eq:intro-trivial}
	\left| N_s - \frac{n}{f(s)}  \right| \le \frac{Q}{2f(s)}.
\end{equation}
In \cite{DudaANS13}, Duda significantly improved upon this by showing that for a variant of \ref{item:duda-second}
\begin{equation} \label{eq:intro-duda}
\left| N_s - \frac{n}{f(s)}  \right| \le \frac{1}{2} \left( \frac{1}{f(s)} + \frac{1}{\min_{t \in \mathcal{S}} f(t)} \right)
\end{equation}
holds. We also mention an algorithm by Dub\'e--Yokoo \cite{dube2019fast} which iteratively tries to improve the allocation based on the dynamics of the streamed variant of asymmetric-numeral-system encoding \cite{DudaANS09}. Unfortunately, for this algorithm, no theoretical guarantees of this type are known.

%In this paper, we introduce a class of algorithms, see Theorem \ref{thm:main-algorithm-criteria} and subsequent remarks, for which we are able to improve upon the bound \eqref{eq:intro-duda}. \todo{based on deadline scheduling} \todo{improve forward reference}

In this paper, we shall introduce a class of algorithms for which we are able to improve upon the bound \eqref{eq:intro-duda}. Before we introduce the class in more detail in the next section, we first give the improved bound in Theorem~\ref{thm:main-discrepancy} and present some derived statements.

\begin{theorem} \label{thm:main-discrepancy}
	For an allocation $A: \mathbb{Z}_{\ge 0} \to \mathcal{S}$ satisfying the criteria of Theorem~\ref{thm:main-algorithm-criteria}, we have that the index $N_s-1$ of the $n$-th occurrence of a symbol $s$ in the allocation $A$ satisfies
	\begin{equation}
		\left| N_s - \frac{n}{f(s)}  \right| \le \frac{1}{f(s)}.		
	\end{equation}
\end{theorem}

We note that this bound can, in general, not be improved further and is thus optimal, cf.\@ Remark~\ref{rem:sharp-discrepancy-bound}. Furthermore, we are able to deduce several important consequences from this bound. First, we show that the natural density of allocations is never far off from the probability distribution~$f$.

\begin{theorem} \label{thm:main-disc-and-KL-bound} Let $A: \mathbb{Z}_{\ge 0} \to \mathcal{S}$ be an allocation satisfying the criteria of Theorem~\ref{thm:main-algorithm-criteria} and $N \in \mathbb{Z}_{> 0}$ a positive integer. Denote by
	\begin{equation} \label{eq:intro-rank-def}
	\rank(s,N-1) =  \bigl| \{n \in \{0,1,\dots,N-1\} \, | \, A[n] = s\} \bigr|
	\end{equation}
	the number of times the symbol $s \in \mathcal{S}$ has been allocated in the first $N$ allocations. Then, the Kullback--Leibler divergence of $\frac{1}{N} \rank(\cdot, N-1)$ relative to $f$ is bounded by
	\begin{equation} \label{eq:main-thm-Kullback-Leibler-divergence}
	D_{\rm KL}\left(f \, \| \, \tfrac{1}{N} \rank(\cdot, N-1) \right) \le \frac{1}{2N^2} \sum_{s \in \mathcal{S}} \frac{1}{f(s)} + O\left(\frac{1}{N^3} \sum_{s \in \mathcal{S}} \frac{1}{f(s)^2} \right).
	\end{equation}
\end{theorem}

The bound \eqref{eq:main-thm-Kullback-Leibler-divergence} on the Kullback--Leibler divergence marks an improvement over \cite[\S 4.2]{DudaANS13}, where the bound
\begin{equation} \label{eq:Duda-KL-divergence}
	\le \frac{1}{2N^2} \sum_{s \in \mathcal{S}} \frac{1}{f(s)} \left( \frac{1}{2} + \frac{f(s)}{2 \min_{t \in S} f(t)}  \right)^2 + O\left( \frac{1}{N^3} \sum_{s \in \mathcal{S}} \frac{1}{f(s)^2} \left( \frac{1}{2} + \frac{f(s)}{2 \min_{t \in S} f(t)}  \right)^3   \right)
\end{equation}
was shown for Duda's algorithm \cite[\S4.1]{DudaANS13}.

As a second consequence of Theorem \ref{thm:main-discrepancy}, we give a bound on the convergence rate of the expected number of bits per symbol in tabled asymmetric numeral systems towards the optimal Shannon entropy \cite{shannon1948mathematical}
\begin{equation} \label{eq:Shannon-entropy}
H = - \sum_{s \in \mathcal{S}} f(s) \log_2(f(s)).
\end{equation}

\begin{theorem} \label{thm:main-entropy}
	Let $A: \mathbb{Z}_{\ge 0} \to \mathcal{S}$ be an allocation satisfying the criteria of Theorem~\ref{thm:main-algorithm-criteria}. Further, let $C: \mathcal{S}^{\ast} \times \mathbb{Z}_{\ge 0} \to \mathbb{Z}_{\ge 0}$ be the tabled-asymmetric-numeral-system encoder based on the allocation $A$ extended to the set of words $\mathcal{S}^{\ast}$ in the alphabet $\mathcal{S}$, see \S \ref{sec:asymmetric-numeral-system} for details. Then, for $m,n \in \mathbb{Z}_{> 0}$, we have that the expected number of bits per symbol when encoding a word of length $m$ to an already encoded message $n$ satisfies
	\begin{multline}
		\frac{1}{m} \sum_{s_1,\dots,s_m \in \mathcal{S}} f(s_1)\cdots f(s_m) \left( {\rm bits}(C(s_1\cdots s_m,n)) - {\rm bits}(n) \right) \\
		= - \sum_{s \in \mathcal{S}} f(s) \log_2(f(s)) + O\left(\frac{1}{m}+\frac{1}{mn} \min\left\{ m, \left(1- \sum_{s \in \mathcal{S}} f(s)^2 \right)^{-1} \right\} \right),
	\end{multline}
	where ${\rm bits}(\ell)= \lfloor \log_2(\ell) \rfloor + 1$, $\ell > 0$, and ${\rm bits}(0)=1$ is the number of bits of a non-negative integer $\ell \in \mathbb{Z}_{\ge 0}$.
\end{theorem}

Once again, Theorem \ref{thm:main-entropy} marks an improvement over what has been shown for the allocation generated by Duda's algorithm \cite[\S4.1]{DudaANS13}. For that allocation, only a bound of
\begin{equation} \label{eq:Duda-tabled-Entropy-diff}
	O\left( \frac{1}{m} + \frac{f(s)}{\min_{t \in S} f(t)} \cdot \frac{1}{mn}  \min\left\{ m, \left(1- \sum_{s \in \mathcal{S}} f(s)^2 \right)^{-1} \right\} \right)
\end{equation}
is known.

\subsection{Class of algorithms}

In order to design algorithms satisfying the conclusion of Theorem \ref{thm:main-discrepancy}, we flip the problem on its head: rather than proving the bound for certain algorithms, we investigate sufficient criteria for an allocation to satisfy the bound. Generalising slightly, we wish to generate allocations, that satisfy
\begin{equation} \label{eq:intro-disc-bnd}
	\left | f(s)N - \rank(s, N-1) \right | \le 1, \quad \forall s \in \mathcal{S}, \forall N \in \mathbb{Z}_{\ge 0},
\end{equation}
where $\rank(s, N-1)$ is as in \eqref{eq:intro-rank-def}. By taking this as a requirement, one is led to a deadline-scheduling problem. More precisely, the problem at hand can be captured in the generalised multiframe task model \cite{baruah1999generalized}, though we do not make further use of this fact. We refer to \cite{stigge2015graph} for a survey on deadline scheduling. 

\begin{table}[!hbp]
	\centering
	\begin{tblr}{
			colspec={Q[c,1em]Q[c,1em]Q[c,1em]Q[c,1em] Q[c,1em]Q[c,1em]Q[c,1em]Q[c,1em] Q[c,1em]Q[c,1em]Q[c,1em]Q[c,1em] Q[c,1em]Q[c,1em]Q[c,1em]},
			rowspec={Q[m,1.4em] Q[m,1.4em]Q[m,1.4em]Q[m,1.4em]Q[m,1.4em]Q[m,1.4em]},
			rowhead = 0,
			rowfoot = 0,
			cell{4}{3,5,8,10,13,15} = {red7},
			cell{4}{4,12} = {azure8},
			cell{3}{8,15} = {azure8},
			cell{3}{5,10} = {green9},
			cell{2}{15} = {green9},
			cell{2}{8} = {Goldenrod},
			cell{1}{15} = {Goldenrod},
			vline{1-16} = {5}{solid},
			vline{1-16} = {6}{solid},
			vline{3-6,8-16} = {4}{solid},
			vline{5-6,8-11,15-16} = {3}{solid},
			vline{8-9,15-16} = {2}{solid},
			vline{15-16} = {1}{solid},
			cell{6}{1,3,5,8,11,13} = {red7},
			cell{6}{2,6,10,12} = {azure8},
			cell{6}{4,9,14} = {green9},
			cell{6}{7,15} = {Goldenrod},
		}
		\cline{15}
		&  &  &  &  &  &  &  &  &  &  &  &  &  & y
		\\ \cline{8,15}
		&  &  &  &  &  &  & y &  &  &  &  &  &  & g
		\\ \cline{5,8,10,15}
		&  &  &  & g &  &  & b &  & g &  &  &  &  & b
		\\ \cline{3-5,8,10,12-13,15}
		&  & r & b & r &  &  & r &  & r &  & b & r &  & r
		\\ \cline{1-15}
		0 & 1 & 2 & 3 & 4 & 5 & 6 & 7 & 8 & 9 & 10 & 11 & 12 & 13 & 14  \\
		\cline{1-15}
		r & b & r & g & r & b & y & r & g & b & r & b & r & g & y  \\
		\cline{1-15}
	\end{tblr}
	%	\vspace{3mm}
	\caption{Deadlines of symbols $\mathcal{S}=\{\text{r}, \text{b}, \text{g}, \text{y}\}$ with probabilities $f(\text{r})=\frac{6}{15}$, $f(\text{b})=\frac{4}{15}$, $f(\text{g})=\frac{3}{15}$, and $f(\text{y})=\frac{2}{15}$ above, together with an earliest-deadline-first allocation table below. Note that at the allocation index $11$, we were not able to allocate the symbol $\text{r}$ as the corresponding job has not been spawned yet.}
	\label{table:deadline-example}
\end{table}

The deadline-scheduling problem is as follows. Let
\begin{equation}
D(s, \ell) = \min\{N \in \mathbb{Z}_{\ge 0}\, | \, f(s)(N+1) \ge \ell \}, \quad \forall s \in \mathcal{S}, \forall \ell \in \mathbb{Z}_{\ge 0}.
\end{equation}
For each symbol $s \in \mathcal{S}$ and every non-negative integer $\ell \in \mathbb{Z}_{\ge 0}$, we spawn an allocation job for the symbol $s$ at the allocation index $D(s, \ell)$ (pre-allocation) with a deadline at the allocation index $D(s, \ell + 1)$. This translates to having the symbol $s$ allocated at least $\ell$ times at allocation indices less than or equal to $D(s, \ell)$ and at most $\ell$ times at allocation indices strictly less than $D(s, \ell)$, for all $s \in \mathcal{S}$ and $\ell \in \mathbb{Z}_{> 0}$.

We prove that this corresponding deadline-scheduling problem is feasible by directly proving that the earliest-deadline-first scheduling algorithm \cite{dertouzos1974control} yields a valid solution. See Table~\ref{table:deadline-example} for an example problem and solution. More generally, we give in Theorem~\ref{thm:main-algorithm-criteria} a flexible criteria on an allocation (algorithm) that is guaranteed to meet all deadlines and thus satisfies the desired bound \eqref{eq:intro-disc-bnd}.

\begin{theorem} \label{thm:main-algorithm-criteria}
	Let $\mathcal{S}$ be finite set symbols with an everywhere positive probability measure $f$ on $\mathcal{S}$ and $A: \mathbb{Z}_{\ge 0} \to \mathcal{S}$ be an allocation of symbols. If for each $N \in \mathbb{Z}_{\ge 0}$,
	
	\begin{enumerate}[label=(\roman*)]
		\item \label{item:intro-criteria-schedulable} $\rank(A[N], N-1) = \lfloor f(A[N])N \rfloor$ or $\lfloor f(A[N])(N+1) \rfloor = \lfloor f(A[N])N \rfloor + 1$; and
		\item \label{item:intro-criteria-foresight} $\forall M \in \mathbb{Z}_{\ge 0}$ with $N+1 \le M <\min\{L \in \mathbb{Z}_{\ge 0} \, | \, \lfloor f(A[N]) L \rfloor \ge \rank(A[N], N) \}$:
		\begin{equation}
			\sum_{s \in \mathcal{S}} \max \left\{0, \lfloor f(s)M \rfloor - \rank(s,N-1) \right\} < M-N
		\end{equation} 
	\end{enumerate}
	hold, then the allocation satisfies the bound \eqref{eq:intro-disc-bnd}.
\end{theorem}

We note that the conditions \ref{item:intro-criteria-schedulable} and \ref{item:intro-criteria-foresight} in Theorem \ref{thm:main-algorithm-criteria} on the $(N+1)$-th allocation $A[N]$ only involve the allocations $A[L]$ for $L < N$. Therefore, the criteria can \emph{a priori} be used as an algorithm template to generate allocations satisfying \eqref{eq:intro-disc-bnd}. Indeed, in \S\ref{sec:discrepancy}, specifically Theorem \ref{thm:bounded-disc-deterministic-algorithm} and Remark \ref{rem:bigger-T-set}, we show that there is always a symbol satisfying the criteria \ref{item:intro-criteria-schedulable} and \ref{item:intro-criteria-foresight}. Hence, such an algorithm is well-defined. Furthermore, we give three explicit algorithms based on this schema in \S\ref{sec:algorithms}.

%We note that whilst the condition \ref{item:intro-criteria-foresight} in Theorem \ref{thm:main-algorithm-criteria} looks like it requires infinite time to verify, it can be verified in finite time. The details of which can be found in \S\ref{sec:discrepancy}. Thus, the criteria can be used for an algorithm to generate allocations satisfying \eqref{eq:intro-disc-bnd}. In \S\ref{sec:algorithms}, we give three such algorithms.

We further remark that Theorem \ref{thm:main-algorithm-criteria} does not make the assumption that the probabilities $f(s)$, $s \in \mathcal{S}$, are rational numbers, in which case an allocation satisfying the conditions of Theorem \ref{thm:main-algorithm-criteria} is not periodic. However, cutting off such an allocation at any point and repeating the pattern indefinitely will give a decent approximation by Theorem \ref{thm:main-disc-and-KL-bound} and can thus be used as tables. Some cut-off points will give a much better approximation. One ought to compare with Dirichlet's approximation theorem. Thus, one might cut off the allocation after $N \ge 1$ allocations where $N$ is such that the quantity $\max_{s\in\mathcal{S}} | f(s)N - \rank(s,N-1)  |$ has reached a new minimum.

\subsection{Overview} In \S \ref{sec:asymmetric-numeral-system}, we give an overview of asymmetric numeral systems. In \S \ref{sec:discrepancy}, we prove the main theorems. In \S \ref{sec:algorithms}, we give concrete examples of algorithms satisfying the criteria of Theorem \ref{thm:main-algorithm-criteria}. In the final sections, \S \ref{sec:evaluation} and \S\ref{sec:discussion}, we compare our algorithms with ones from the literature and give some concluding remarks.

\subsection{Acknowledgements}
We would like to extend our gratitude to Albert-Jan Yzelman for his continued support and guidance. We also thank Toni B\"ohnlein and P\'al Andr\'as Papp for discussions on this and surrounding topics. We are also grateful to Jarek Duda for discussions on an earlier version of the manuscript.

%Last but not least, we would like to thank our employer the Huawei Computing Lab for a fruitious work environment.

RS would like to thank JentGent for a wonderful introductory video on compression \cite{JentGentCompression} that got him interested in the topic.
\section{Asymmetric numeral systems} \label{sec:asymmetric-numeral-system}

\subsection{General}

In this section, we go over some of the variants of asymmetric numeral systems \cite{DudaANS09,DudaANS13}.

Let $\mathcal{S}$ be a non-empty finite set of symbols and $f$ an everywhere positive probability measure on $\mathcal{S}$, where $f(s)$ is the probability of the new symbol being $s$. Given an allocation $A: \mathbb{Z}_{\ge 0} \to \mathcal{S}$, we define two functions
\begin{equation}
\begin{aligned}
\rank: \mathcal{S} \times \mathbb{Z}_{\ge 0} &\to \mathbb{Z}_{> 0} \\
(s, n) & \mapsto \bigl | \{m \in \mathbb{Z}_{\ge 0} \, | \, A[m]=s \text{ and } m \le n \} \bigr |,
\end{aligned}
\end{equation}
which maps $(s, n)$ to the number of occurrences of the symbol $s$ in the allocation $A$ up to and including $n$, and
\begin{equation}
\begin{aligned}
\select: \mathcal{S} \times \mathbb{Z}_{>0} &\to \mathbb{Z}_{\ge 0} \\ 
(s, n) & \mapsto \min \{m \in \mathbb{Z}_{\ge 0} \, | \, \rank(s, m) \ge n \},
\end{aligned}
\end{equation}
which maps $(s, n)$ to the index of the $n$-th occurrence of the symbol $s$ in the allocation $A$.

We can now define the asymmetric-numeral-system encoding scheme
\begin{equation} \label{eq:ANS-encoding}
\begin{aligned}
C: \mathcal{S} \times \mathbb{Z}_{\ge 0} &\to \mathbb{Z}_{\ge 0} \\
(s, n) &\mapsto 
\select(s, n+1)
\end{aligned}
\end{equation}
with corresponding decoding scheme
\begin{equation} \label{eq:ANS-decoding}
\begin{aligned}
D: \mathbb{Z}_{\ge 0} &\to \mathcal{S} \times \mathbb{Z}_{\ge 0} \\ 
n &\mapsto (A[n], \rank(A[n], n) - 1).
\end{aligned}
\end{equation}

Furthermore, we naturally extend the encoding scheme $C$ to $\mathcal{S}^{\ast} \times \mathbb{Z}_{\ge 0}$, where $\mathcal{S}^{\ast}$ is the free monoid generated by $\mathcal{S}$ (words in the alphabet $\mathcal{S}$), by setting $C(\eps, n) = n$ for the neutral element (empty word) $\eps\in\mathcal{S}^{\ast}$ and $C(sw, n) = C(s, C(w, n))$ for $s \in \mathcal{S}$ and $w \in \mathcal{S}^{\ast}$.

\subsubsection{Expected number of bits}

The expected number of bits to encode a symbol from an initial state $n$ when using the asymmetric-numeral-system encoding scheme \eqref{eq:ANS-encoding} to encode words $w \in \mathcal{S}^{\ast}$ of length $m$ is
\begin{equation} \label{eq:expected-number-bits}
{\rm EB}(n) = \frac{1}{m} \sum_{s_1,\dots,s_m \in \mathcal{S}} f(s_1)\cdots f(s_m) \left( {\rm bits}(C(s_1\cdots s_m,n)) - {\rm bits}(n) \right),
\end{equation}
where %${\rm bits}(\ell)$ denotes the number of bits of a non-negative integer $\ell \in \mathbb{Z}_{\ge 0}$.
\begin{equation} \label{eq:number-of-bits}
\begin{aligned} 
	{\rm bits}: \mathbb{Z}_{\ge 0} &\to \mathbb{Z}_{>0} \\
	n &\mapsto \begin{cases} 1, & \text{if } n=0, \\ \lfloor \log_2(n) \rfloor + 1 & \text{otherwise}, \end{cases}
\end{aligned}
\end{equation}
denotes the number of bits of a non-negative integer $\ell \in \mathbb{Z}_{\ge 0}$. %
This should be compared to the optimal Shannon entropy \eqref{eq:Shannon-entropy}.

Since ${\rm bits}(C(s,n)) - {\rm bits}(n) \approx -\log_2(n / C(s,n))$, we require the allocation of symbols to be made in such a way that $n \slash C(s, n)$ is close to $f(s)$ for all symbols $s\in \mathcal{S}$ and states $n\in \mathbb{Z}_{\ge 0}$ in order to be close to the optimal. Indeed, we have the following proposition which bounds the difference of the expected number of bits for words and the Shannon entropy \eqref{eq:Shannon-entropy}.

\begin{proposition} \label{prop:entropy-convergence}
	Let the discrepancy
	\begin{equation} \label{eq:sup-discrepancy}
	D = \sup_{N \in \mathbb{Z}_{> 0}} \max_{s \in \mathcal{S}} \left | Nf(s) - \rank(s, N-1) \right |.
	\end{equation}
	Then, for $m,n \in \mathbb{Z}_{> 0}$, we have
	\begin{equation} \label{eq:ANS-discrepancy-bound}
	{\rm EB}(n)
	=  - \sum_{s \in \mathcal{S}} f(s) \log_2(f(s)) + O\left( \frac{1}{m} + \frac{D+1}{mn} \min\left\{ m , \left( 1 - \sum_{s \in \mathcal{S}} f(s)^2 \right)^{-1}  \right\} \right).
	\end{equation}
\end{proposition}

\begin{proof}
	First, we may replace ${\rm bits}(\ell) = \log_2(\ell) + O(1)$. This introduces an error of $O(\frac{1}{m})$. Secondly, from the definition of the discrepancy \eqref{eq:sup-discrepancy} and the encoding \eqref{eq:ANS-encoding}, we have
	\begin{equation}
	C(s,\ell) = \frac{\ell + 1}{f(s)} - 1 + O\left( \frac{D}{f(s)} \right).
	\end{equation}
	Hence, it follows that
%	\begin{equation}
%	C(s_1 \cdots s_m, n) = \frac{n+1}{f(s_1)\cdots f(s_m)} - 1 + O\left(  \frac{mD}{f(s_1)\cdots f(s_m)}  \right)
%	\end{equation}
	\begin{equation}
	C(s_1 \cdots s_m, n) = \frac{n+1}{f(s_1)\cdots f(s_m)} - 1 + O\left( \sum_{i = 1}^{m} \frac{D}{f(s_1)\cdots f(s_i)}  \right)
	\end{equation}
	and furthermore
%	\begin{equation}
%	\log_2(C(s_1 \cdots s_m, n)) = \log_2\left( \frac{n}{f(s_1)\cdots f(s_m)} \right) + O\left( \frac{1+mD}{n}   \right).
%	\end{equation}
	\begin{equation}
	\log_2(C(s_1 \cdots s_m, n)) = \log_2\left( \frac{n}{f(s_1)\cdots f(s_m)} \right) + O\left( \frac{D+1}{n}\sum_{i=0}^m f(s_{i+1})\cdots f(s_m)    \right).
	\end{equation}
	Thus, 
%	\begin{multline}
%	\frac{1}{m}\sum_{\substack{s_1\dots s_m \in \mathcal{S}^{\ast} \\ s_i \in \mathcal{S}}} f(s_1)\cdots f(s_m) \left( \log_2(C(s_1 \cdots s_m, n)) - \log_2(n) \right) \\
%	=  - \frac{1}{m} \sum_{\substack{s_1\dots s_m \in \mathcal{S}^{\ast} \\ s_i \in \mathcal{S}}} f(s_1)\cdots f(s_m) \log_2(f(s_1) \cdots f(s_m)) + O\left( \frac{1}{m} + \frac{D}{n} \right) \\
%	= - \sum_{s \in \mathcal{S}} f(s) \log_2(f(s)) + O\left( \frac{1}{m} + \frac{D}{n}  \right).
%	\end{multline}
	\begin{equation} \begin{aligned}
	&\frac{1}{m}\sum_{\substack{s_1, \dots, s_m \in \mathcal{S} }} f(s_1)\cdots f(s_m) \left( \log_2(C(s_1 \cdots s_m, n)) - \log_2(n) \right) \\
	=&  - \frac{1}{m} \sum_{\substack{s_1, \dots, s_m \in \mathcal{S}}} f(s_1)\cdots f(s_m) \log_2(f(s_1) \cdots f(s_m)) + O\left( \frac{1}{m} + \frac{D+1}{mn}\sum_{i=0}^m\left(\sum_{s \in \mathcal{S}} f(s)^2 \right)^i \right) \\
	=& - \sum_{s \in \mathcal{S}} f(s) \log_2(f(s)) + O\left( \frac{1}{m} + \frac{D+1}{mn} \min\left\{ m, \left( 1- \sum_{s \in \mathcal{S}} f(s)^2 \right)^{-1} \right\}  \right).
	\end{aligned} \end{equation}
	This concludes the proof.
\end{proof}

We note that Theorem \ref{thm:main-entropy} is a simple corollary of Proposition \ref{prop:entropy-convergence} and Theorem \ref{thm:main-algorithm-criteria}.

\subsection{Tabled asymmetric numeral system} \label{sec:tabled-asymmetric-numeral-system}

In tabled asymmetric numeral systems the allocation $A : \mathbb{Z}_{\ge 0} \to \mathcal{S}$ is assumed to be periodic. This significantly improves the memory requirement and computational efficiency of encoding and decoding. Besides our new algorithms, see \S\ref{sec:discrepancy}-\ref{sec:algorithms}, we briefly introduce three algorithms from the literature. Two of which, we shall discuss in this section and one in the next. 

We assume that $f(\mathcal{S}) \subseteq \mathbb{Q}$ and that $Q \in \mathbb{Z}_{>0}$ is such that $f(\mathcal{S})Q \subseteq \mathbb{Z}$. Often such a $Q$ is taken to be minimal, but this is not necessary.

\subsubsection{Ranged table} \label{sec:range-table}
In \cite[\S2.3]{DudaANS13}, the symbols $\mathcal{S} = \{s_1, s_2, \dots, s_t \}$ are ordered descendingly according to $f(s)$, $s \in \mathcal{S}$, that is $f(s_1) \ge f(s_2) \ge \dots \ge f(s_t)$. The allocation of symbols is then the periodic extension of the following table:
\begin{equation}
\underbrace{s_1,s_1,\dots,s_1}_{f(s_1)Q\text{-times}}, \underbrace{s_2,s_2,\dots,s_2}_{f(s_2)Q\text{-times}}, \dots,
\underbrace{s_t,s_t,\dots,s_t}_{f(s_t)Q\text{-times}}.
\end{equation}

\subsubsection{Duda's table} \label{sec:Duda-table}
In \cite[\S4]{DudaANS13}, the following algorithm was given, see Algorithm \ref{alg:Duda-table}. There, the minimum is taken in lexicographical order.

\begin{algorithm}[!h]
	\DontPrintSemicolon
	\SetNlSty{textsc}{}{}
	\SetAlgoNlRelativeSize{-1}
	\caption{Duda's allocation \label{alg:Duda-table}}
	\KwData{A finite set of symbols $\mathcal{S}$ and an everywhere positive probability measure $f$ on $\mathcal{S}$.}
	\KwResult{An allocation of symbols $A: \mathbb{Z}_{\ge 0} \to \mathcal{S}$.}
	\BlankLine
	$\mathcal{Q}  \leftarrow \left\{\left(\frac{1}{2f(s)}, f(s), s \right) \in \mathbb{R} \times \mathbb{R} \times \mathcal{S} \, | \, s \in \mathcal{S} \right\}$ \; \label{line:duda-init}
	\For{$N=0,1,2,3,\dots$}{
		$(v, \rule{2.4mm}{.5pt}, s) \leftarrow {\rm extractMin}(\mathcal{Q})$ \;
		$\mathcal{Q}.{\rm add}\left(\left(v + \frac{1}{f(s)}, f(s), s\right)\right)$ \;
		$A[N] \leftarrow s$\;
	}
	\Return $A$\;
\end{algorithm}

\begin{remark}
	In Duda's initial paper \cite[\S5]{DudaANS09}, the heap $\mathcal{Q}$ in Line \ref{line:duda-init} was initialised with $(\frac{1}{f(s)}, f(s), s)$ instead.
\end{remark}

\subsection{Stream asymmetric numeral system} \label{sec:stream-asymmetric-numeral-system}

Stream asymmetric numeral system is a variant of the tabled encoding scheme that aims to combat the everchanging encoded data during encoding. It does so by renormalising the encoded data to a fixed interval whilst keeping track of the renormalising in a (bit) stream.

Let $B \in \mathbb{Z}$ be an integer with $B \ge 2$, which will be used in the renormalisation process. For computer architecture reasons, a power of $2$ is often chosen as $B$. Fix an interval $\mathcal{I}$ of integers of the shape
\begin{equation} \label{eq:stream-shape-interval-gen}
\mathcal{I} = [M , M+1, \dots, BM-1],
\end{equation}
for some $M \in \mathbb{Z}_{> 0}$. For simplicity and efficiency, we shall assume throughout that $Q \mid M$ such that each symbol appears in $\mathcal{I}$ with the correct frequency. Furthermore, we define for each symbol $s \in \mathcal{S}$ the preimage under the encoding $C(s, \cdot)$, %see Equation \eqref{eq:ANS-encoding},
\begin{equation} \label{eq:stream-shape-interval-symbol}
\mathcal{I}_s \coloneqq \{n \in \mathbb{Z}_{\ge 0} \, | \, C(s, n) \in \mathcal{I}\} = [Mf(s), Mf(s)+1, \dots, BMf(s) - 1].
\end{equation}
Since $Q \mid M$, we have $Mf(s) \in \mathbb{Z}_{>0}$. Hence, $\mathcal{I}_s$ is of the same shape \eqref{eq:stream-shape-interval-gen} (with different $M$). Intervals of said shape are called $B$-absorbing and satisfy the properties:
\begin{enumerate}[label=(\roman*)]
\item $m \ge bM \Rightarrow \exists! \ell \in \mathbb{Z}_{>0}: \lfloor m \slash B^{\ell} \rfloor \in \mathcal{I}$, and
\item $m \le M - 1 \Rightarrow \forall (b_n)_{n > 0} \subseteq \{0,1,\dots,B-1\}: \exists! \ell \in \mathbb{Z}_{>0}: nB^{\ell}+b_1B^{\ell -1} + \dots + b_{\ell} \in \mathcal{I}$.
\end{enumerate}

Before we can discuss the streamed version, we need to introduce the stream. The stream $\mathcal{B}^{\ast}$ is the free monoid generated by $\{0,1,\dots,B-1\}$. In other words, $\mathcal{B}^{\ast}$ is the set of words written with the alphabet $\{0,1,\dots,B-1\}$ with concatenation. Streamed asymmetric-numeral-system encoding is then given by

\begin{equation} \label{eq:stream-ANS-encoding}
\begin{aligned}
\overrightarrow{C}: \mathcal{S} \times \mathcal{I} \times \mathcal{B}^{\ast} &\to \mathcal{I} \times \mathcal{B}^{\ast} \\
(s, n, \vect{b}) &\mapsto \begin{cases} (C(s, n), \vect{b}), & \text{if } n \in \mathcal{I}_s, \\
%\overrightarrow{C}(s, \lfloor n \slash B \rfloor, (n \ \mod(B))\vect{b}), & \text{otherwise}, \\
(C(s, n'), b_{1}\cdots b_{\ell}\vect{b}), & \begin{aligned}[c] &\text{if } n=n'B^{\ell} + b_{1}B^{\ell -1} + \dots + b_{\ell} \\ &\text{with } b_i \in \{0,1,\dots,B-1\} \\ &\text{and } n'=\lfloor n \slash B^{\ell} \rfloor \in \mathcal{I}_s. \end{aligned} 
\end{cases}
\end{aligned}
\end{equation}
The corresponding decoding scheme is given as
\begin{equation} \label{eq:stream-ANS-decoding}
\begin{aligned}
\overrightarrow{D}: \mathcal{I} \times \mathcal{B}^{\ast} &\to (\mathcal{S} \times \mathcal{I} \times \mathcal{B}^{\ast}) \cup \{\text{Error}\} \\ 
(n, \vect{b}) &\mapsto \begin{cases} (s, m, \vect{b}), & \text{if } D(n)=(s,m) \text{ and } m \in \mathcal{I}, \\
(s, m', \vect{b'}) & \begin{aligned}[c] &\text{if } D(n)=(s, m), \vect{b}=b_1\cdots b_{\ell} \vect{b'} \text{ with }b_i \in \{0,1,\dots, B-1\}, \\ &\text{and } m'= mB^{\ell}+b_1B^{\ell -1}+\dots + b_{\ell} \in \mathcal{I}, \end{aligned} \\
\text{Error},  & \text{otherwise}.
\end{cases}
\end{aligned}
\end{equation}

\subsubsection{Expected word length increase per symbol} \label{sec:stream-exp-word-length-per-symbol}

By leaving out the stream $\mathcal{B}^{\ast}$ in the stream encoding \eqref{eq:stream-ANS-encoding}, we get a time-homogeneous Markov chain on $\mathcal{I}$. The Perron--Frobenius theorem tells us then that there is at least one invariant probability distribution on $\mathcal{I}$. We note that multiple invariant probability distributions may indeed occur, see \cite[\S 6.1]{DudaANS09} and \cite{fujisaki2020irreducibility} for a discussion on the irreducibility of the Markov chain. Nevertheless, given an invariant probability distribution  $p$ on $\mathcal{I}$, we can compute the expected word length increase per symbol. It is given by

\begin{equation} \label{eq:expected-word-length-stream}
	{\rm EWL} = \sum_{s \in \mathcal{S}} f(s) \left( (\lambda_s - 1) \sum_{n = M}^{Mf(s)B^{\lambda_s} - 1} p(n) + \lambda_s \sum_{n = Mf(s)B^{\lambda_s}}^{BM-1} p(n) \right),
\end{equation}
where $\lambda_s = \lceil -\log_B(f(s))  \rceil$, see \cite[Eq.\@ (1)]{dube2019fast}. This should be compared with the Shannon entropy \eqref{eq:Shannon-entropy} with logarithm base $B$.

\subsubsection{Dub\'e--Yokoo sort-based table}
\label{sec:dube-yokoo-table}

The finite nature of the stream variant slightly distorts the probability distribution compared to the asymptotic nature of the tabled variant. In order to take this into account, Dub\'e--Yokoo \cite[\S III]{dube2019fast} have developed an iterative algorithm.

Beginning with the ranged table allocation, see \S\ref{sec:range-table}, they compute an associated invariant probability measure\footnote{They did not specify which one to take in case there are multiple invariant probability measures.}. From here, they compute a permutation which orders descendingly the invariant probabilities and apply it to the allocation table to get a new allocation table. This process is repeated until a loop is detected. Finally, they take the allocation with the smallest expected word length per symbol \eqref{eq:expected-word-length-stream} out of the computed allocations.

\section{Discrepancy}\label{sec:discrepancy}

%In this section, we describe our algorithm to generate tables for asymmetric numeral systems and prove that they achieve the optimal bound. We have written this section in a self-contained manner since the results may be of interest to other disciplines, for example scheduling synchronous data-flows \cite{discrepancy-synchronous-data-flow}.

In this section, we describe our algorithm to generate tables for asymmetric numeral systems and prove that they achieve the optimal bound. We have written this section in a self-contained manner since the results are of interest to other disciplines, for example scheduling theory, see \cite{stigge2015graph} for a survey. Indeed, the algorithm presented here may be seen as a more general version of an earliest-deadline-first scheduling algorithm \cite{dertouzos1974control} in the generalised multiframe task model \cite{baruah1999generalized}.

Let $V$ be a non-empty finite set and $f:\mathcal{S} \to \, ]0,1]$ a map such that $\sum_{s \in \mathcal{S}} f(s) = 1$, which represents the desired natural density of occurrences of the elements in $\mathcal{S}$ in a sequence. Note that we allow the natural densities to be real numbers and not just rational.

Let $\vect{x} = (x_n)_{n \ge 0} \subseteq \mathcal{S}$ be a sequence. We define the discrepancy for $s \in \mathcal{S}$ and $N \in \mathbb{Z}_{\ge 0}$ as
\begin{equation} \label{eq:discrepancy}
	D(s, N) = D(s,N;\vect{x},\mathcal{S}) =f(s)N-\sum_{\substack{n < N \\ x_n = s}} 1.
\end{equation}
It measures the discrepancy (difference) between number of expected occurrences and actual occurrences of a symbol. The aim is thus to keep the discrepancy small in absolute value. We shall abbreviate the number of occurrences of $s$ in the sequence $\vect{x}$ up to and excluding the $N$-th element as $n(s,N)$, that is
\begin{equation} \label{eq:disc-seq-occurrences}
n(s,N) = \sum_{\substack{n < N \\ x_n = s}} 1.
\end{equation}

We begin with a trivial bound on the discrepancy for periodic sequences.

\begin{proposition} \label{prop:disc-trivial-bound}
	Let $\mathcal{S}$ be a non-empty finite set and $\vect{x} = (x_n)_{n \ge 0} \subseteq \mathcal{S}$ a periodic sequence of period length $Q$. Suppose further that $f: \mathcal{S} \to \, ]0,1]$ denotes the natural density of each $s\in \mathcal{S}$, that is $f(s) = \frac{1}{Q} \bigl | \{n \in \{0,1,\dots,Q-1\} \, | \, x_n = s\} \bigr |$. Then,
	\begin{equation}\label{eq:disc-trivial-bound}
		\sup_{\substack{s \in \mathcal{S} \\ N \in \mathbb{Z}_{\ge 0}}} \left|  D(s,N) \right| \le \tfrac{1}{2}Q.
	\end{equation}
\end{proposition}
\begin{proof}
	The discrepancy $D(s,N)$ is minimised for $N=Qf(s)$ and $x_0=x_1=\dots=x_{Qf(s)-1}=s$. In this case, we have
	\begin{equation}
	0 \ge D(s,Qf(s)) = f(s)Qf(s)-Qf(s) = Qf(s)(f(s)-1) \ge - \tfrac{1}{2}Q.
	\end{equation}
	Likewise, we have that the discrepancy $D(s,N)$ is maximised for $N=Q(1-f(s))$ and $x_n \neq s$ for $n = 0,1,\dots,Q(1-f(s))-1$. In this case, we have
	\begin{equation}
	0 \le D(s,Qf(s)) = f(s)Q(1-f(s)) \le \tfrac{1}{2}Q.
	\end{equation}
\end{proof}

We now describe a class of algorithms and show that they significantly improve upon the trivial bound. In fact, the bound we give is optimal, see Remark \ref{rem:sharp-discrepancy-bound}.

\begin{theorem} \label{thm:bounded-disc-deterministic-algorithm}
	Let $\mathcal{S}$ be a non-empty finite set and $f: \mathcal{S} \to \, ]0,1]$ such that $\sum_{s \in \mathcal{S}} f(s) = 1$. Define a sequence $\vect{x} = (x_n)_{n \ge 0} \subseteq \mathcal{S}$ recursively where the $(N+1)$-th symbol $x_N$ is chosen from the set of symbols $\mathcal{T}(N) \subseteq \mathcal{S}$ for $N \in \mathbb{Z}_{\ge 0}$, where
	\begin{equation} \label{eq:disc-T-def}
	\begin{aligned}
	\mathcal{T}(N) = \Bigl\{  t & \in \mathcal{S}  \  \big| \ \bigl( n(t,N)=\lfloor f(t) N \rfloor \text{ or } \lfloor f(t) (N+1) \rfloor = \lfloor f(t) N \rfloor + 1 \bigr) \text{ and } \\
	&\forall M \in \mathbb{Z}_{\ge 0}, M > N:  \sum_{\substack{s \in \mathcal{S} }} \max\left\{0, \lfloor f(s)M \rfloor - n(s,N) - \delta_{s,t}  \right\}  \le M - N - 1 \Bigr\},
	\end{aligned}
	\end{equation}
	and $\delta_{s,t}$ is the Kronecker delta. Then, the sequence is well-defined, meaning $x_N$ can always be chosen (equivalently $\mathcal{T}(N) \neq \emptyset$ for all $N \in \mathbb{Z}_{\ge 0}$), and the following bound on the discrepancy holds
	\begin{equation} \label{eq:discrepancy-bound-above}
	\sup_{\substack{s \in \mathcal{S} \\ N \in \mathbb{Z}_{\ge 0}}} \left|  D(s,N) \right| \le 1. 
	\end{equation}
\end{theorem}

\begin{remark} \label{rem:sharp-discrepancy-bound}
	We note that the bound \eqref{eq:discrepancy-bound-above} is sharp as the example $f \equiv \rquotient{1}{|\mathcal{S}|}$ with $|\mathcal{S}| \to \infty$ shows.
\end{remark}

%\begin{remark} \todo{turn into a proposition}
%	If $f(\mathcal{S}) \subseteq \mathbb{Q}$, then the sequence generated is necessarily periodic. Indeed, let $Q \in \mathbb{Z}_{>0}$ be such that $f(s)Q \in \mathbb{Z}, \forall s \in \mathcal{S}$. Then,
%	\begin{equation}
%	Q = \sum_{s \in S} n(s,Q) \ge \sum_{s \in \mathcal{S}} \lfloor f(s)Q \rfloor = \sum_{s \in \mathcal{S}} f(s)Q = Q.
%	\end{equation}
%	Hence, $n(s,Q)=f(s)Q, \forall s \in \mathcal{S}$. From here, it is a straightforward induction to show that $x_{N+Q}=x_{N}$ for $N \in \mathbb{Z}_{\ge 0}$ (if ties in the minimum are resolved in the same way). 
%\end{remark}

\begin{proof}[Proof of Theorem \ref{thm:bounded-disc-deterministic-algorithm}]
 We will prove inductively over $N$ that
 \begin{enumerate}[label=(\roman*)]
 	\item \label{item:disc-ineq} $\forall N \in \mathbb{Z}_{\ge 0}: \forall s \in \mathcal{S}$: $n(s,N) \in \left\{ \lfloor f(s)N \rfloor , \lfloor f(s)N \rfloor + 1 \right\}$,
 	
 	\item \label{item:disc-non-empty} $\forall N \in \mathbb{Z}_{\ge 0}:$ the set $\left\{ s \in \mathcal{S} \mid n(s,N) = \lfloor f(s)N \rfloor  \right\}$ is non-empty,
 	
 	\item \label{item:disc-foresight} $\forall M, N \in \mathbb{Z}_{\ge 0}$ such that $M \ge N$:
 	\begin{equation} \label{eq:disc-foresight-ineq}
 		\sum_{\substack{s \in \mathcal{S} }} \max\left\{0, \lfloor f(s)M \rfloor - n(s,N)  \right\}  \le M - N,
 	\end{equation}
 	and,
 	\item \label{item:T-non-empty}
 	$\forall N \in \mathbb{Z}_{\ge 0}: \mathcal{T}(N) \neq \emptyset$.
 \end{enumerate}
	We note that \ref{item:disc-ineq} ensures that \eqref{eq:discrepancy-bound-above} is satisfied, \ref{item:disc-non-empty} and \ref{item:T-non-empty} makes sure that the sequence $\vect{x}$ is well-defined, and \ref{item:disc-foresight} guarantees that all future deadlines induced by the condition \eqref{eq:discrepancy-bound-above} can be met.

 	For $N=0$, we have $n(s,0) = 0$ for all $v \in V$, thus \ref{item:disc-ineq} and \ref{item:disc-non-empty} are trivial in this case. Furthermore, we have
 	\begin{equation}
 		\sum_{s \in \mathcal{S}} \lfloor f(s) M \rfloor \le \sum_{s \in \mathcal{S}} f(s) M = M,
 	\end{equation}
 	which shows \ref{item:disc-foresight} for $N=0$.
 	
 	In general, we claim that \ref{item:disc-ineq} implies \ref{item:disc-non-empty}. Assume $n(s,N) \neq \lfloor f(s) N \rfloor $ for all $s \in \mathcal{S}$. Then, by \ref{item:disc-ineq}, we have
 	\begin{equation}
 	N = \sum_{s \in \mathcal{S}} n(s,N) = \sum_{s \in \mathcal{S}} \left( \lfloor f(s) N \rfloor + 1  \right) > \sum_{s \in \mathcal{S}} f(s) N = N,
 	\end{equation}
 	a contradiction! Hence, \ref{item:disc-non-empty} holds.
 	
 	Next, we claim that \ref{item:disc-ineq}, \ref{item:disc-non-empty}, and \ref{item:disc-foresight} together imply \ref{item:T-non-empty}. We do so by showing, that 
 	\begin{equation} \label{eq:t_N-ast-def}
 		t^{\ast}_N = \argmin_{\substack{s \in \mathcal{S} \\ n(s,N) = \lfloor f(s) N \rfloor }} % \text{ or} \\ \lfloor f(s) (N+1) \rfloor = \lfloor f(s) N \rfloor + 1}}
 		\min\left\{M \in \mathbb{Z}_{\ge 0} \mid \lfloor f(s)M \rfloor \ge n(s,N) + 1  \right\} \in \mathcal{T}(N).
 	\end{equation}
 	We first note that $t_N^{\ast}$ exist by \ref{item:disc-non-empty}. Let
 	\begin{equation}
 		M_N^{\ast} = \min\left\{M \in \mathbb{Z}_{\ge 0} \mid \lfloor f(t_N^{\ast})M \rfloor \ge n(t_N^{\ast},N) + 1  \right\}.
 	\end{equation}
 	We know that $M_N^{\ast} \ge N + 1$ since $n(t_N^{\ast}, N) \ge  \lfloor f(t_N^{\ast})N \rfloor$ by \ref{item:disc-ineq}. Further, let
 	\begin{equation} \label{eq:M_N-def}
 		M_N = \min \left\{ M \in \mathbb{Z}_{\ge 0} \cup \{\infty\} \ \bigg| \ \sum_{s \in \mathcal{S}} \max\left\{0, \lfloor f(s)M \rfloor - n(s,N)  \right\}  = M - N > 0 \right\}.
 	\end{equation}
 	We claim that
 	\begin{equation} \label{eq:M_N-M_N-ast-ineq}
 		M_N \ge M_N^{\ast}.
 	\end{equation}
 	If $M_N = \infty$, there is nothing to prove. Otherwise, we prove Inequality \eqref{eq:M_N-M_N-ast-ineq} by showing that there is an $s \in \mathcal{S}$ such that
 	\begin{equation} \label{eq:optimal-proof-existence-vertex-flex}
 		\lfloor f(s)M_N \rfloor - n(s,N) \ge 1 \quad \text{ and } \quad n(s,N) = \lfloor f(s) N \rfloor. 
 	\end{equation}
 	If this were not the case, we would have from \ref{item:disc-ineq} that
 	\begin{equation}
 		\lfloor f(s)M_N \rfloor - n(s,N) \ge 1 \quad \Rightarrow \quad n(s,N) = \lfloor f(s) N \rfloor + 1. 
 	\end{equation}
 	Let $\mathcal{U} \subseteq \mathcal{S}$, be the subset of $s$'s for which $\lfloor f(s)M_N \rfloor - n(s,N) \ge 1$ holds. Then,
 	\begin{equation}\begin{aligned}
 		M_N-N &= \sum_{\substack{s \in \mathcal{U} }} \max\left\{ 0, \lfloor f(s)M_N \rfloor - n(s,N) \right\}  \\
 		&= \sum_{\substack{s \in \mathcal{U} }} \left( \lfloor f(s)M_N \rfloor - \lfloor f(s)N) \rfloor - 1 \right) \\
 		&< \sum_{s \in \mathcal{U}} f(s) (M_N-N)  \\
 		&\le \sum_{s \in \mathcal{S}} f(s) (M_N-N) = M_N-N,
 	\end{aligned}\end{equation}
 	which is a contradiction due to the strict inequality in the third line! Hence, there is a $s \in \mathcal{S}$ such that \eqref{eq:optimal-proof-existence-vertex-flex} holds. In particular, Inequality \eqref{eq:M_N-M_N-ast-ineq} holds.
 	
 	Let now $M \in \mathbb{Z}_{\ge 0}$ with $M > N$. If $M < M_N^{\ast} \le M_N$, we have
 	\begin{align}
 	\label{eq:disc-chosen-diff-small}
 	\max\left\{ 0, \lfloor f(t^{\ast}_N)M \rfloor - n(t^{\ast}_N,N) - 1\right\} 
 	&\le \max\left\{ 0, \lfloor f(t^{\ast}_N)M \rfloor - n(t^{\ast}_N,N) \right\}, \\
 	\label{eq:disc-induction-diff-small}
 	\sum_{\substack{s \in \mathcal{S} }} \max\left\{ 0, \lfloor f(s)M \rfloor - n(s,N) \right\}
 	&\le M-N-1. 
 	\end{align}
 	On the other hand, if $M \ge M_N^{\ast}$, we have from \ref{item:disc-foresight} that
 	\begin{align}
 	\label{eq:disc-chosen-diff-large}
 	\max\left\{ 0, \lfloor f(t^{\ast}_N)M \rfloor - n(t^{\ast}_N,N) - 1\right\} 
 	&\le \max\left\{ 0, \lfloor f(t^{\ast}_N)M \rfloor - n(t^{\ast}_N,N) \right\} - 1, \\
 	\label{eq:disc-induction-diff-large}
 	\sum_{\substack{s \in \mathcal{S} }} \max\left\{ 0, \lfloor f(s)M \rfloor - n(s,N) \right\}
 	&\le M-N. 
 	\end{align}
 	In both cases, summing the two inequalities yields
 	\begin{equation}
 		\sum_{\substack{s \in \mathcal{S} }} \max\left\{0, \lfloor f(s)M \rfloor - n(s,N) - \delta_{s,t_N^{\ast}}  \right\}  \le M - N - 1.
 	\end{equation}
 	Hence, $t_N^{\ast} \in \mathcal{T}(N)$, which proves \ref{item:T-non-empty}. 	
 	
 	For $N \ge 1$, the claim \ref{item:disc-ineq} follows from \ref{item:disc-foresight} with $N-1$. Indeed, since $\lfloor f(s) M \rfloor $ from $M=N-1$ to $M=N$ can only increase by at most one, \ref{item:disc-ineq} for $N$ is satisfied for those $s \in \mathcal{S}$ for which either $n(s,N-1) = \lfloor f(s) (N-1) \rfloor + 1$ or $\lfloor f(s) (N-1) \rfloor = \lfloor f(s) N \rfloor$. From \ref{item:disc-foresight} with $N-1$ and $M=N$, we have
 	\begin{equation}
 	\sum_{\substack{s \in \mathcal{S} \\ n(s,N-1) = \lfloor f(s) (N-1) \rfloor \\ \lfloor f(s) N \rfloor \ge \lfloor f(s) (N-1) \rfloor +1 }} 1 \le \sum_{s \in \mathcal{S}} \max\left\{ 0, \lfloor f(s)N \rfloor -n(s,N-1 ) \right\} \le 1 .
 	\end{equation}
 	Thus, either the remaining set of $s$'s is empty, in which case there is nothing left to prove, or there is at most one such $s$, but then $\mathcal{T}(N-1) = \{s\}$ by Definition \eqref{eq:disc-T-def}, hence $n(s,N) = n(s,N-1)+1 = \lfloor f(s) (N-1) \rfloor + 1 = \lfloor f(s) N \rfloor$. This concludes \ref{item:disc-ineq}.

 	Likewise, for $N \ge 1$, the claim \ref{item:disc-foresight} for $N$ follows trivially from \ref{item:T-non-empty} with $N - 1$ as the definition of the set $\mathcal{T}(N-1)$ \eqref{eq:disc-T-def} implies the desired inequality. This concludes the proof.
\end{proof}

\begin{remark} \label{rem:bigger-T-set}
	The proof of Theorem \ref{thm:bounded-disc-deterministic-algorithm} more generally shows that
	\begin{multline} \label{eq:T-extended-criteria}
%		\begin{aligned}
		\Bigl\{  s  \in \mathcal{S}  \  \big| \ \bigl( n(s,N)=\lfloor f(s) N \rfloor \text{ or } \lfloor f(s) (N+1) \rfloor = \lfloor f(s) N \rfloor + 1 \bigr) \\ \text{and }
		\lfloor f(s) M_N \rfloor \ge n(s,N) + 1 \Bigr\} \subseteq \mathcal{T}(N),
%		\end{aligned}
	\end{multline}
	where $M_N$ is defined as in \eqref{eq:M_N-def}. This concludes Theorem \ref{thm:main-algorithm-criteria}.
\end{remark}

\begin{corollary}[Theorem \ref{thm:main-discrepancy}] \label{cor:nth-occurrence}
	Let $(x_n)_{n \ge 0} \subseteq \mathcal{S}$ be a sequence generated by an algorithm in Theorem \ref{thm:bounded-disc-deterministic-algorithm}. Then, the $n$-th occurrence of an element $s \in \mathcal{S}$ occurs at $x_{N-1}$, where $N$ satisfies
	\begin{equation} \label{eq:nth-occurrence}
	\left| N -\frac{n}{f(s)} \right| \le \frac{1}{f(s)}.
	\end{equation}
\end{corollary}

\begin{proof}
	This follows immediately from Inequality \eqref{eq:discrepancy-bound-above}.
\end{proof}

\begin{corollary}[Theorem \ref{thm:main-disc-and-KL-bound}] \label{cor:Kullback-Leibler-divergence}
	Let $(x_n)_{n \ge 0} \subseteq \mathcal{S}$ be a sequence generated by an algorithm in Theorem \ref{thm:bounded-disc-deterministic-algorithm} and $N \in \mathbb{Z}_{>0}$ be a positive integer. Define a probability measure $q_N$ on $\mathcal{S}$ by
	\begin{equation}
	q_N(s) = \tfrac{1}{N}n(s,N) = \tfrac{1}{N} \bigl | \left\{n \in \{0,1,\dots,N-1\} \, | \, x_n = s \right\} \bigr |.
	\end{equation}
	Then, the Kullback--Leibler divergence of $q_N$ relative to $f$ satisfies
	\begin{equation} \label{eq:Kullback-Leibler-divergence}
		D_{\rm KL}(f \, \| \, q_N) \le \frac{1}{2N^2} \sum_{s \in \mathcal{S}} \frac{1}{f(s)} + O\left(\frac{1}{N^3} \sum_{s \in \mathcal{S}} \frac{1}{f(s)^2} \right).
	\end{equation}
\end{corollary}

\begin{proof}
	We have
	\begin{equation}
%	\begin{aligned}
	D_{\rm KL}(f \, \| \, q_N) %= \sum_{s \in \mathcal{S}} f(s) \log\left( \frac{f(s)}{q_N(s)} \right) 
	= \sum_{s \in \mathcal{S}} f(s) \log\left( \frac{Nf(s)}{n(s,N)} \right) 
	= -\sum_{s \in \mathcal{S}} f(s) \log\left( 1+ \frac{n(s,N)-Nf(s)}{Nf(s)} \right).
%	= - \sum_{s \in \mathcal{S}} f(s) \left[\left( \frac{n(s,N)-Nf(s)}{Nf(s)} \right) - \frac{1}{2} \left( \frac{n(s,N)-Nf(s)}{Nf(s)} \right)^2 + O\left( \left( \frac{n(s,N)-Nf(s)}{Nf(s)} \right)^3 \right)  \right]
%	\end{aligned}
	\end{equation}
	From here, we conclude the corollary after substituting the second order Taylor expansion of $\log(1+x)=x-\frac{1}{2}x^2+O(x^3)$ and using the bound \eqref{eq:discrepancy-bound-above} on the discrepancy.
%	\begin{equation}
%	D_{\rm KL}(f \, \| \, q_n) \le \frac{1}{2N^2} \sum_{s \in \mathcal{S}} \frac{1}{f(s)} + O\left(\frac{1}{N^3} \sum_{s \in \mathcal{S}} \frac{1}{f(s)^2} \right).
%	\end{equation}
\end{proof}

Finally, we prove that if $f(\mathcal{S}) \subseteq \mathbb{Q}$, then the sequence generated by Theorem \ref{thm:bounded-disc-deterministic-algorithm} is necessarily periodic.

\begin{proposition} \label{prop:periodic-rational}
	Let $\vect{x}=(x_n)_{n \ge 0} \subseteq \mathcal{S}$ be a sequence generated by an algorithm in Theorem \ref{thm:bounded-disc-deterministic-algorithm}. If there is a $Q \in \mathbb{Z}_{>0}$ such that $f(s)Q \in \mathbb{Z}, \forall s \in \mathcal{S}$, then $x_{N+Q}=x_{N}$ for $N \in \mathbb{Z}_{\ge 0}$.
\end{proposition}

\begin{proof}
	From the proof of Theorem \ref{thm:bounded-disc-deterministic-algorithm}, specifically \ref{item:disc-ineq}, we have
	\begin{equation}
		Q = \sum_{s \in S} n(s,Q) \ge \sum_{s \in \mathcal{S}} \lfloor f(s)Q \rfloor = \sum_{s \in \mathcal{S}} f(s)Q = Q.
	\end{equation}
	Hence, $n(s,Q)=f(s)Q, \forall s \in \mathcal{S}$. From here, it is a straightforward induction to show that $x_{N+Q}=x_{N}$ for $N \in \mathbb{Z}_{\ge 0}$ (if the choices are made in the same way).
\end{proof}

%The bound \eqref{eq:discrepancy-bound-above} on the discrepancy implies that the $n$-th occurrence of an element $s \in \mathcal{S}$ in the sequence generated by the algorithm in Proposition \ref{prop:bounded-disc-deterministic-algorithm} occurs at $x_{N-1}$, where $N$ satisfies
%\begin{equation} \label{eq:nth-occurrence}
%\left| N -\frac{n}{f(s)} \right| \le \frac{1}{f(s)}.
%\end{equation}
%
%
%This is an improvement over the bound for the algorithm in \S\ref{sec:Duda-table} from \cite{DudaANS13}. There, the bound
%\begin{equation} \label{eq:nth-occurrence-duda}
%\left| N -\frac{n}{f(s)} \right| \le \frac{1}{2} \left( \frac{1}{f(s)}+\frac{1}{\min_{s \in \mathcal{S}} f(s) } \right).
%\end{equation}
%was shown.
\section{Algorithms}
\label{sec:algorithms}

In this section, we lay out several explicit algorithms based on Theorem \ref{thm:bounded-disc-deterministic-algorithm}.

\begin{algorithm}[!h]
	\DontPrintSemicolon
	\SetNlSty{textsc}{}{}
	\SetAlgoNlRelativeSize{-1}
	\caption{Earliest deadline first \label{alg:table-allocation-earliest-deadline-first}}
	\KwData{A finite set of symbols $\mathcal{S}$ and an everywhere positive probability measure $f$ on $\mathcal{S}$.}
	\KwResult{An allocation of symbols $A: \mathbb{Z}_{\ge 0} \to \mathcal{S}$.}
	\BlankLine
	${\rm numAlloc}[s] \leftarrow 0 $, \textit{for all} $s\in \mathcal{S}$ \;
	\For{$N=0,1,2,3,\dots$}{
		%		$\mathcal{T} = \{s \in \mathcal{S} \, | \, {\rm numAlloc}[s] = \lfloor f(s)N  \rfloor \}$ \;
		$U, V \leftarrow +\infty$ \;
		\For{$s \in \mathcal{S}$}{
			\If{${\rm numAlloc}[s] \neq \lfloor f(s)N \rfloor$}{
				\Continue
			}
			\If{$\lfloor f(s) U \rfloor \ge {\rm numAlloc}[s] + 1 $}{
				$U' \leftarrow \min \{M \in \mathbb{Z}_{\ge 0} \, | \, \lfloor f(s)M \rfloor \ge {\rm numAlloc}[s] +1 \}$ \tcp*{$U' \le U$}
				$V' \leftarrow {\rm numAlloc}[s] - f(s)(N+1)$\;
				\If{$U' < U$ \Or $V' < V$ \Or $(V'=V$ \And $f(A[N]) \ge f(s))$}{
					$A[N] \leftarrow s$ \;
					$U \leftarrow U'$ \;
					$V \leftarrow V'$\;
				}
			}
		}
		${\rm numAlloc}[A[N]] \leftarrow {\rm numAlloc}[A[N]] + 1 $ \;
	}
	\Return $A$\;
\end{algorithm}

Our first algorithm, Algorithm \ref{alg:table-allocation-earliest-deadline-first}, is perhaps the most straightforward one as it is directly based on the proof of Theorem \ref{thm:bounded-disc-deterministic-algorithm}. It employs the strategy of allocating the earliest deadline first \cite{dertouzos1974control} together with a more elaborate tie-breaking rule which aims to minimise the discrepancy.

\begin{algorithm}[!h]
	\DontPrintSemicolon
	\SetNlSty{textsc}{}{}
	\SetAlgoNlRelativeSize{-1}
	\caption{Shifted priorities \label{alg:table-allocation-shifted-priorities}}
	\KwData{A finite set of symbols $\mathcal{S}$, an everywhere positive probability measure $f$ on $\mathcal{S}$, and an integer $Q \in \mathbb{Z}_{>0}$ such that $f(\mathcal{S})Q \subseteq \mathbb{Z}$.}
	\KwResult{An allocation of symbols $A: \mathbb{Z}_{\ge 0} \to \mathcal{S}$.}
	\BlankLine
	${\rm numAlloc}[s] \leftarrow 0, \, \forall s\in \mathcal{S}$ \;
	${\rm deadline} \leftarrow [(\min \{ M \in \mathbb{Z}_{\ge 0} | \lfloor f(s)M \rfloor \ge n \}, s) \PythFor s \in \mathcal{S} \PythFor n = 1,\dots,Qf(s) ]$ \;
	${\rm spawnQ} \leftarrow {\rm deadline}.{\rm copy}()$\;
	${\rm symbolQ} \leftarrow [(\min \{ M \in \mathbb{Z}_{\ge 0} | \lfloor f(s)M \rfloor \ge 1 \}, f(s), s) \PythFor s \in \mathcal{S}]$ \label{line:shifted-prio-initial-spawns} \;
	${\rm spawnCntr}[s] \leftarrow 1, \ \forall s \in \mathcal{S}$\;
	\For{$N=0,1,2,\dots, Q-1$ \label{line:shifted-prio-for-spawns-begin}}{
		\If{$N \le Q-1$}{
			\For{$(R, s) \in {\rm spawnQ}$ \St $R = N$}{
				${\rm symbolQ}.{\rm add}((\min \{ M \in \mathbb{Z}_{\ge 0} \, |\, \lfloor f(s)M + \tfrac{1}{2} \rfloor \ge {\rm spawnCntr}[s]+1 \}, f(s), s))$\;
				${\rm spawnCntr}[s] \leftarrow {\rm spawnCntr}[s] + 1$ \label{line:shifted-prio-for-spawns-end}\;
			}
		}
		$(\_, \_, t) \leftarrow {\rm symbolQ}.{\rm min}()$\;
		$M \leftarrow {\rm deadline}.{\rm getEarliestDeadlineOfSymbol}(t)$\;
		$L \leftarrow N+1$ \label{line:greedy-disc-meet-deadline-begin}\;
		\While{$L < M$ \And $| \{(K, s) \in {\rm deadline} \, | \, K \le L \} | < L - N$}{
			$L \leftarrow L + 1$\;
		}
		\If{$L < M$}{
			$(\_, \_, t) \leftarrow \min \{(K, f(s), s) \in {\rm symbolQ} \, | \, \exists (P,s) \in {\rm deadline} \text{ with } P \le L \}$ \label{line:greedy-disc-meet-deadline-end} \;
		}
		$A[N] \leftarrow t$\;
		${\rm deadline}.{\rm removeSmallestWithSymbol}(A[N])$ \;
		${\rm symbolQ}.{\rm removeSmallestWithSymbol}(A[N])$ \;
		${\rm numAlloc}[A[N]] \leftarrow {\rm numAlloc}[A[N]] + 1 $ \;
	}
	\Return $A$\;
\end{algorithm}

\begin{algorithm}[!h]
	\DontPrintSemicolon
	\SetNlSty{textsc}{}{}
	\SetAlgoNlRelativeSize{-1}
	\caption{Greedy discrepancy minimisation \label{alg:table-Greedy-discrepancy-minimisation}}
	\KwData{A finite set of symbols $\mathcal{S}$, an everywhere positive probability measure $f$ on $\mathcal{S}$, and an integer $Q \in \mathbb{Z}_{>0}$ such that $f(\mathcal{S})Q \subseteq \mathbb{Z}$.}
	\KwResult{An allocation of symbols $A: \mathbb{Z}_{\ge 0} \to \mathcal{S}$.}
	\BlankLine
	${\rm numAlloc}[s] \leftarrow 0, \, \forall s\in \mathcal{S}$ \;
	${\rm deadline} \leftarrow [(\min \{ M \in \mathbb{Z}_{\ge 0} | \lfloor f(s)M \rfloor \ge n \}, s) \PythFor s \in \mathcal{S} \PythFor n = 1,\dots,Qf(s) ]$ \;
	${\rm spawnQ} \leftarrow {\rm deadline}.{\rm copy}()$\;
	${\rm symbolMultiSet} \leftarrow \mathcal{S}$ \label{line:greedy-disc-initial-spawn} \;
	\For{$N=0,1,2,\dots, Q-1$}{
		\If{$N \le Q-1$}{
			\For{$(R, s) \in {\rm spawnQ}$ \St $R = N$}{
				${\rm symbolMultiSet}.{\rm add}(s)$ \label{line:greedy-disc-for-spawn-end}\;
			}
		}
		$U \leftarrow -\infty$\;
		$L \leftarrow N + 1$ \label{line:greedy-disc-meet-deadline-criteria-begin}\;
		\While{
			$L < Q$ \And $| \{(K, s) \in {\rm deadline} \, | \, K \le L \} | < L - N$
		}{
			$L \leftarrow L + 1$\;
		}
		\For{$s \in {\rm symbolMultiSet}$}{
			\If{${\rm deadline}.{\rm getEarliestDeadlineOfSymbol}(s) \le L$ \label{line:greedy-disc-meet-deadline-criteria-end}}{
				$U' \leftarrow f(s)(N+1)-{\rm numAlloc}[s]$\;
				\If{$U' > U$ \Or $(U' = U$ \And $f(s) < f(A[N]))$}{
					$U \leftarrow U'$\;
					$A[N] \leftarrow s$\;
				}
			}
		}
		${\rm deadline}.{\rm removeSmallestWithSymbol}(A[N])$ \;
		${\rm symbolMultiSet}.{\rm remove}(A[N])$ \;
		${\rm numAlloc}[A[N]] \leftarrow {\rm numAlloc}[A[N]] + 1 $ \;
	}
	\Return $A$\;
\end{algorithm}

Algorithms \ref{alg:table-allocation-shifted-priorities} and \ref{alg:table-Greedy-discrepancy-minimisation} are more advanced as they make use of the extended characterisation given by Remark \ref{rem:bigger-T-set} of allowed symbols to allocate. This extra foresight allows for sufficient flexibility to incorporate a symbol allocation priority and exert some control over when a particular symbol should be allocated. It has already been observed by Duda \cite{DudaANS13, DudaANS21} that in streamed asymmetric numeral systems low probability symbols $s\in \mathcal{S}$ are best allocated around (or more precisely just before) the points $(n + \frac{1}{2})/f(s)$ for $n = 0, 1, \dots, Qf(s) - 1$, where $f(s)$ is the probability of the symbol $s$ and $Q$ is the table length. Thus, Algorithms \ref{alg:table-allocation-shifted-priorities} and \ref{alg:table-Greedy-discrepancy-minimisation} use their flexibility to do exactly that.
Algorithm \ref{alg:table-allocation-shifted-priorities} accomplished this in a similar fashion as Duda's algorithm \ref{alg:Duda-table}, namely by incorporating a soft deadline for when each symbol desires to be allocated.
Algorithm \ref{alg:table-Greedy-discrepancy-minimisation}, on the other hand, achieves the desired outcome by prioritising minimising the discrepancy such that the number of allocation of a symbol $s \in \mathcal{S}$ in the first $N$ allocations is most likely the nearest integer to the number of expected allocations $f(s)N$.

\begin{remark}
	We note that when the number of symbols $|\mathcal{S}|$ is two, Algorithm~\ref{alg:table-Greedy-discrepancy-minimisation} behaves similarly to asymmetric binary system of Duda \cite[\S 3]{DudaANS09}. If we let $\mathcal{S}=\{a,b\}$, then  Algorithm~\ref{alg:table-Greedy-discrepancy-minimisation}, puts $A[N]=a$ if and only if $f(a)(N+1)-{\rm numAlloc}[a] \ge \frac{1}{2}$, whereas (one of the mentioned variants of) asymmetric binary systems puts $A[N]=a$ if and only if $f(a)(N+1)-{\rm numAlloc}[a] \ge 1$.
\end{remark}

We now briefly explain how the algorithms match the template of Theorem \ref{thm:bounded-disc-deterministic-algorithm}. Algorithm \ref{alg:table-allocation-earliest-deadline-first} simply allocates $t_N^{\ast}$, see Equation \eqref{eq:t_N-ast-def}, at each step. In Algorithm \ref{alg:table-allocation-shifted-priorities}, the ${\rm spawnQ}$ makes sure that ${\rm symbolQ}$ only contains those symbols $s \in \mathcal{S}$ for which $n(s,N)= \lfloor f(s)N \rfloor$ or $\lfloor f(s)(N+1) \rfloor =\lfloor f(s)N \rfloor + 1$ at every step $N$. This occurs in Lines \ref{line:shifted-prio-initial-spawns} and \ref{line:shifted-prio-for-spawns-begin}-\ref{line:shifted-prio-for-spawns-end}. Lines \ref{line:greedy-disc-meet-deadline-begin}-\ref{line:greedy-disc-meet-deadline-end} are responsible for ensuring that the to be allocated symbol satisfies the second criteria in \eqref{eq:T-extended-criteria}.
Algorithm \ref{alg:table-Greedy-discrepancy-minimisation} works similarly. In Lines \ref{line:greedy-disc-initial-spawn}-\ref{line:greedy-disc-for-spawn-end}, ${\rm spawnQ}$ guarantees at every step $N$ that ${\rm symbolMultiSet}$ only contains symbols $s \in \mathcal{S}$ for which $n(s,N)= \lfloor f(s)N \rfloor$ or $\lfloor f(s)(N+1) \rfloor =\lfloor f(s)N \rfloor + 1$. Lines \ref{line:greedy-disc-meet-deadline-criteria-begin}-\ref{line:greedy-disc-meet-deadline-criteria-end} assure that the allocated symbol satisfies the second criteria in \eqref{eq:T-extended-criteria}.

We note that Algorithm \ref{alg:table-allocation-earliest-deadline-first}, could also be made more efficient with the spawning of allocatable symbols employed in Algorithms \ref{alg:table-allocation-shifted-priorities} and \ref{alg:table-Greedy-discrepancy-minimisation}, though this would only lead to runtime improvements if a simpler and more efficient tie-breaking rule were to be used.

Finally, we mention that whilst Algorithms \ref{alg:table-allocation-shifted-priorities} and \ref{alg:table-Greedy-discrepancy-minimisation} as presented assume rational symbol probabilities, they can be adapted to deal with the irrational case as well.

\section{Evaluation} \label{sec:evaluation}

In this section, we compare the algorithms from \S \ref{sec:algorithms} against the various algorithms from the literature for tabled and streamed asymmetrical numeral systems, see \S\ref{sec:range-table}, \S\ref{sec:Duda-table}, and \S\ref{sec:dube-yokoo-table}. Recall the notation from \S\ref{sec:asymmetric-numeral-system}.

\subsection{Symbol probabilities} \label{sec:eval-samples}

We examine five varied symbol probabilities in detail. The samples we use are in Table \ref{table:sample-set}. To get the probabilities from the occurrences, the displayed values have to be divided by their sum.

\begin{table}[htbp]
	\centering
	\begin{tblr}{
			colspec = { Q[l,m] | Q[c,m, wd=20em]},
			row{1} = {font=\bfseries},
			rowhead = 1,
			rowfoot = 0,
		}
		Sample & Occurrences of symbols  \\ \hline \hline
		Linear & $1,2,3,4,5,6,7,8$ \\ \hline
		Fibonacci & $1,1,2,3,5,8,13,21$ \\ \hline
		Uniform random & $5, 6, 10, 10, 12, 17, 17, 18$ \\ \hline
		Zipfian random & $1, 1, 1, 1, 2, 5, 5, 14$ \\ \hline
		Alphabet & $82, %a
		15, %b
		28, %c
		43, %d
		127, %e
		22, %f
		20, %g
		61, %h
		70, %i
		2, %j
		8, %k
		40, %l
		24,$ %m
		
		$67, %n
		75, %o
		19, %p
		1, %q
		60, %r
		63, %s
		91, %t
		28, %u
		10, %v
		24, %w
		2, %x
		20, %y
		1% %z
		$
	\end{tblr}
%	\vspace{3mm}
	\caption{The linear sample consists of the first eight positive integers, the Fibonacci sample consists of the first eight Fibonacci numbers, the uniform random sample was drawn uniformly at random from the interval $[1, 20]$, the Zipfian random sample was drawn randomly from the interval $[1, 20]$ with respect to the Zipfian distribution (truncated at $20$), and the alphabet sample is an approximation to the frequencies of letters in the English alphabet taken from \cite[Table 1.1]{lewand2000cryptological}.}
	\label{table:sample-set}
\end{table}

%\begin{enumerate}[label=(\roman*)]
%%	\item $[2,5,10]$, \label{item:data-duda}
%	\item Linear $[$1,2,3,4,5,6,7,8,9,10$]$, \label{item:data-linear}
%	\item Fibonacci $[$1,1,2,3,5,8,13,21,34,55$]$, \label{item:data-fibonacci}
%	\item Uniform random $[$12, 29, 30, 31, 48, 51, 51, 54, 65, 80$]$, \label{item:data-uniform-random}
%	\item Zipfian random $[$1, 1, 2, 4, 5, 9, 13, 25, 46, 64$]$, and \label{item:data-zipfian-random}
%	\item Alphabet $[$82, %a
%			 15, %b
%			 28, %c
%			 43, %d
%			127, %e
%			 22, %f
%			 20, %g
%			 61, %h
%			 70, %i
%			  2, %j
%			  8, %k
%			 40, %l
%			 24, %m
%			 67, %n
%			 75, %o
%			 19, %p
%			  1, %q
%			 60, %r
%			 63, %s
%			 91, %t
%			 28, %u
%			 10, %v
%			 24, %w
%			  2, %x
%			 20, %y
%			  1% %z
%	$]$. \label{item:data-alphabet}
%%	\item $[$0.08167, %a
%%			0.01492, %b
%%			0.02782, %c
%%			0.04253, %d
%%			0.12702, %e
%%			0.02228, %f
%%			0.02015, %g
%%			0.06094, %h
%%			0.06966, %i
%%			0.00153, %j
%%			0.00772, %k
%%			0.04025, %l
%%			0.02406, %m
%%			0.06749, %n
%%			0.07507, %o
%%			0.01929, %p
%%			0.00095, %q
%%			0.05987, %r
%%			0.06327, %s
%%			0.09056, %t
%%			0.02758, %u
%%			0.00978, %v
%%			0.02360, %w
%%			0.00150, %x
%%			0.01974, %y
%%			0.00074% %z
%%			$]$. \label{item:data-alphabet}
%\end{enumerate}

For the performance plots, we used uniform and Zipfian random samples, one hundred of each kind. For the uniform random samples, eight integers were drawn uniformly at random from the interval $[1,20]$. For the Zipfian random samples, again eight integers from the interval $[1, 20]$ were drawn with respect to the Zipfian distribution (truncated at $20$). If the computation of the expected word length increase per symbol \eqref{eq:expected-word-length-stream} failed due to numerical instability in the computation of the invariant probability measure, the sample was thrown out and replaced with a different one.

\subsection{Tabled asymmetric numeral systems}

We are unable to properly compare the expected number of bits per symbol \eqref{eq:expected-number-bits} as the word length increases. This is due to the exponential growth of number of computations compared to the linear convergence rate. In addition, for small word length, the expected number of bits heavily depends on the initial value making a fair comparison difficult. Instead, we tweak the encoding scheme \eqref{eq:ANS-encoding} a bit:
\begin{equation} \label{eq:table-encoding-adapted}
	\begin{aligned}
		\widetilde{C} : \mathcal{S} \times \mathbb{Z}_{\ge 0} & \to \mathbb{Z}_{> 0} \\
		(s, n) & \mapsto \select(s, n+1) + 1.
	\end{aligned}
\end{equation}

\begin{figure}[hbtp]
	\centering
	\includegraphics[scale=0.36]{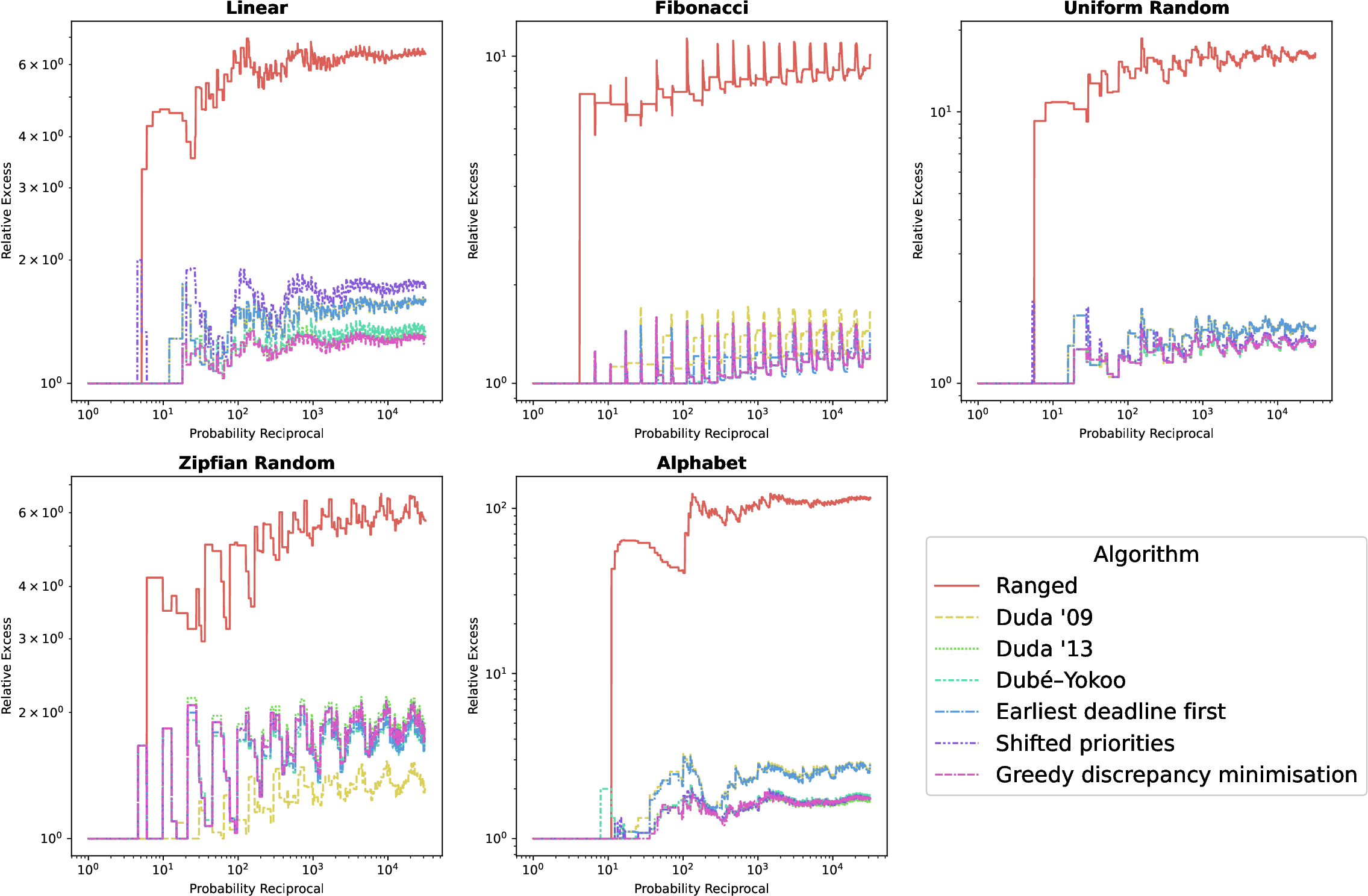}
	\caption{Relative excess in table asymmetric numeral systems plotted against the reciprocal of the probability.}
	\label{fig:table-entropy-loss}
\end{figure}

%\begin{equation} \label{eq:table-decoding-adapted}
%\begin{aligned}
%\widetilde{D} : \mathbb{Z}_{> 0} & \to \mathcal{S} \times \mathbb{Z}_{\ge 0} \\
%n & \mapsto (A[n-1], \rank(A[n-1], n-1) - 1) 
%\end{aligned}
%\end{equation}
This tweak, or more precisely its extension to $\mathcal{S}^{\ast} \times \mathbb{Z}_{\ge 0}$, allows one to define a bijection from $\mathcal{S}^{\ast}$ to $\mathbb{Z}_{\ge 0}$ mapping $w \in \mathcal{S}^{\ast}$ to $\widetilde{C}(w, 0)$.
%
%\begin{equation} \label{eq:words-integer-bijection}
%\begin{aligned}
%\widetilde{\iota} : \mathcal{S}^{\ast} & \to \mathbb{Z}_{\ge 0} \\
%w & \mapsto \widetilde{C}(w, 0).
%\end{aligned}
%\end{equation}
%
For a probability $p \in ]0,1[$, we can then define the set
\begin{equation}
\mathcal{S}^{\ast}_p = \{w \in \mathcal{S}^{\ast} \, | \, f(s_1)\cdots f(s_m) \ge p, \text{ where } w=s_1\cdots s_m, \text{ with } s_i \in \mathcal{S}, \forall i=1,\dots,m  \}
\end{equation}
consisting of words whose product of probabilities of the symbols it is made out of is at least $p$. These words can in some sense be seen as the most likely ones and should therefore have the smallest encodings. We thus define the relative excess as
\begin{equation} \label{eq:relative-excess}
	{\rm RE}(p) = \frac{1}{|\mathcal{S}^{\ast}_p|}\max_{w \in \mathcal{S}^{\ast}_p} \widetilde{C}(w,0),
\end{equation}
which we present in Figure \ref{fig:table-entropy-loss}.

\subsection{Stream asymmetric numeral systems}
\begin{figure}[hbtp]
	\centering
	\includegraphics[scale=0.36]{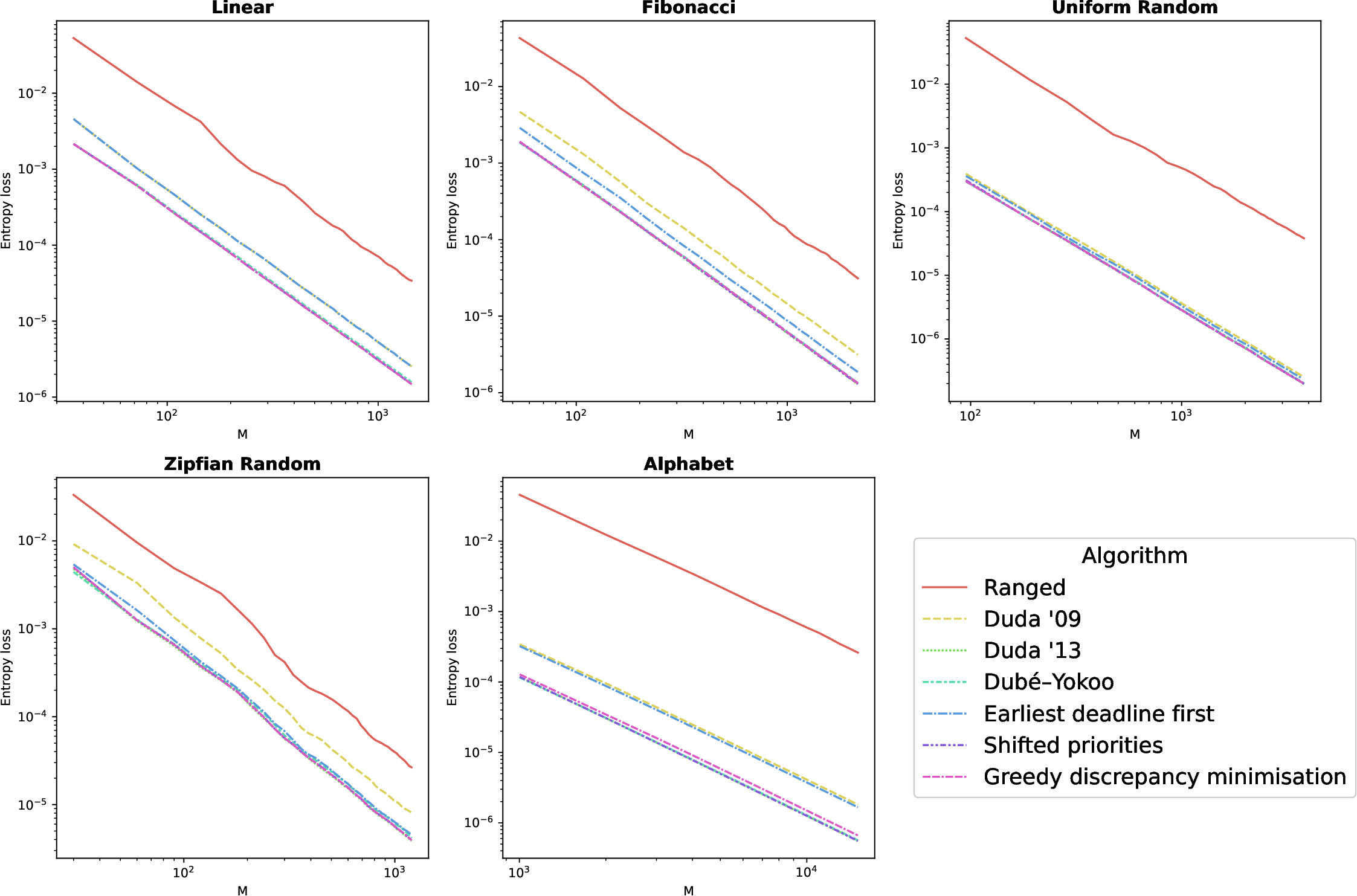}
	\caption{Entropy loss in stream asymmetric numeral systems plotted against the beginning/length $M$ of the interval $\mathcal{I}$, see Equation \eqref{eq:stream-shape-interval-gen}, for the samples in Table \ref{table:sample-set}.}
	\label{fig:stream-entropy-loss}
\end{figure}

In a first step, we compute the entropy loss as the difference between expected word length increase per symbol ${\rm EWL}$ using the basis $B=2$, see Equation \eqref{eq:expected-word-length-stream}, and the Shannon entropy \eqref{eq:Shannon-entropy} for the samples in Table \ref{table:sample-set}. The results are found in Figure \ref{fig:stream-entropy-loss}.

\begin{figure}[hbtp]
	\centering
	\includegraphics[scale=0.36]{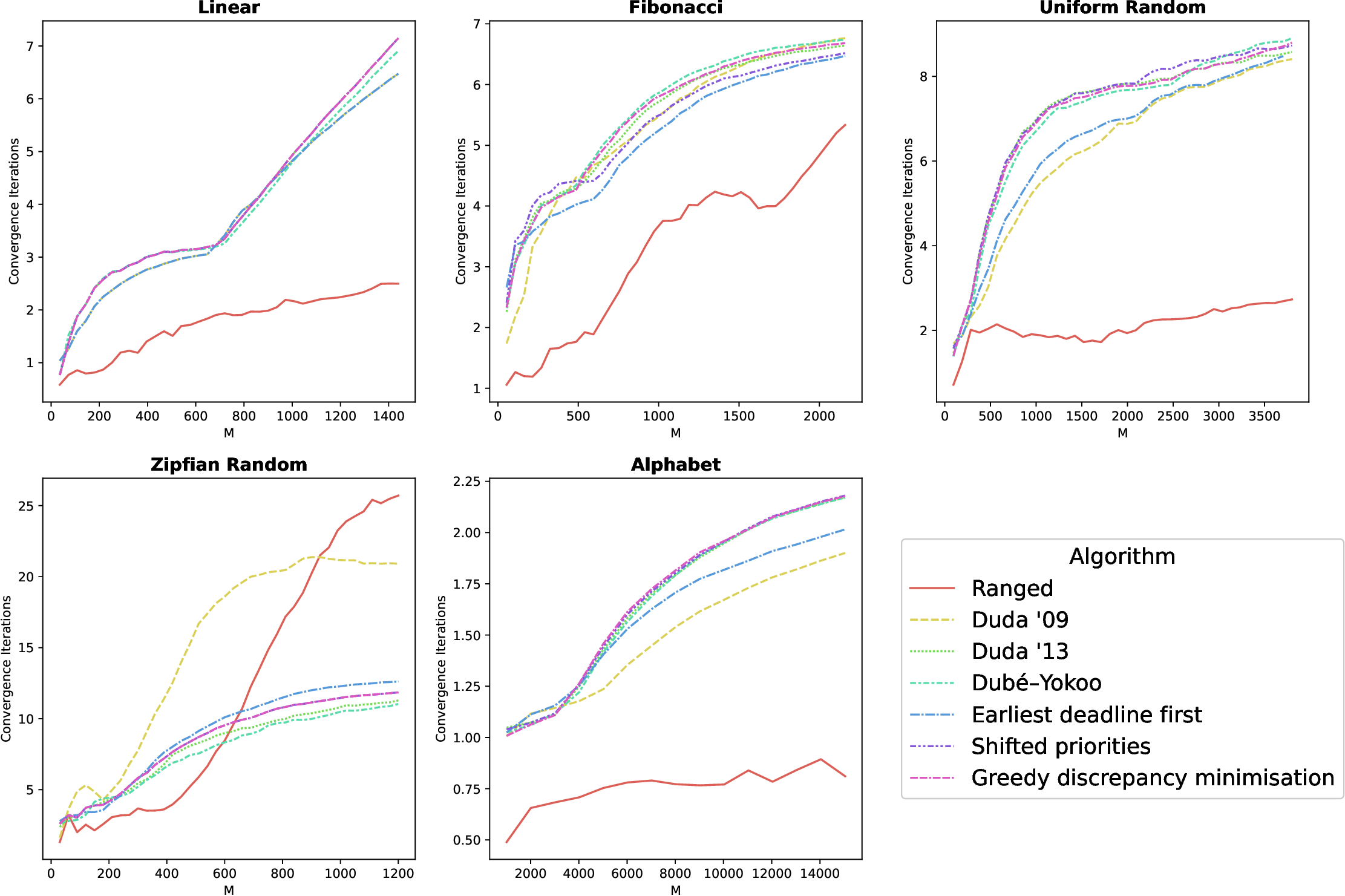}
	\caption{Order of magnitude of the number of iterations required for convergence to the invariant measure in stream asymmetric numeral systems plotted against the beginning/length $M$ of the interval $\mathcal{I}$, see Equation \eqref{eq:stream-shape-interval-gen}, for the samples in Table \ref{table:sample-set}.}
	\label{fig:stream-EV}
\end{figure}

In the second figure, Figure \ref{fig:stream-EV}, we display the scale of number of iterations (encodings of symbols) needed for a probability measure on $\mathcal{I}$ to start to converge towards the invariant probability measure (if unique). More precisely, we have computed this as the logarithm of the ratio of the largest eigenvalue, which is one, and the second largest eigenvalue in absolute value. This often overlooked statistic is important as it tells us how quickly one can expect the increase in the (bit) stream to converge to the computed expected word length increase per symbol ${\rm EWL}$ from Figure \ref{fig:stream-entropy-loss}.

\begin{figure}[hbtp]
	\centering
	\includegraphics[scale=0.36]{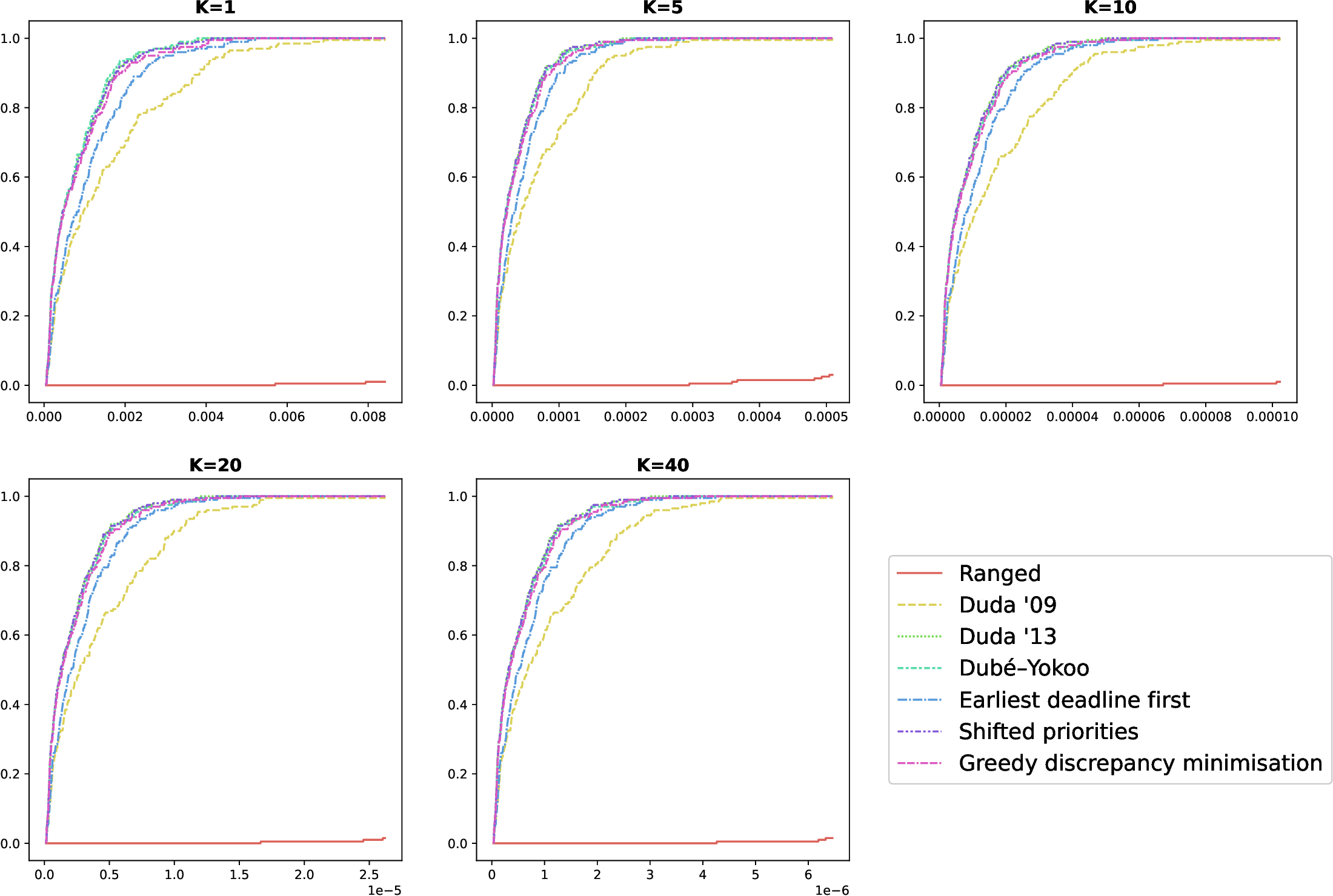}
	\caption{Performance profiles on one hundred uniform random and one hundred Zipfian random samples. Here, the beginning/length of the interval $\mathcal{I}$ in streamed asymmetric numeral systems, see Equation \eqref{eq:stream-shape-interval-gen}, is $M=K\cdot Q$, where $Q$ is the length of the table. On the x-axis, we have the threshold for the quotient of the entropy loss and Shannon entropy, and on the y-axis, we have the relative amount of samples satisfying the threshold.}
	\label{fig:stream-Performance}
\end{figure}

In the last figure, Figure \ref{fig:stream-Performance}, we show the performance in entropy loss of the algorithms in a performance profile \cite{dolan2002benchmarking} on a large set of randomly generated samples, see \S\ref{sec:eval-samples}.

\section{Discussion}
\label{sec:discussion}

In the evaluation section, \S\ref{sec:evaluation}, in particular the performance profiles in Figure \ref{fig:stream-Performance}, we see that Algorithm \ref{alg:table-allocation-earliest-deadline-first}, earliest deadline first, performs better than the earlier algorithm of Duda \cite[\S5]{DudaANS09} but falls short of all other algorithms (excluding ranged). This shows that whilst discrepancy is very important in the macro scale it fails to fully capture the nuances in the micro scale, which are of significance in streamed asymmetric numeral systems. 
We thus believe it to be important to better understand and understand quantitatively the heuristics of Dub\'e--Yokoo \cite[\S III]{dube2019fast} and Duda \cite[\S 4]{DudaANS13} in streamed asymmetric numeral systems. For example, Duda's heuristic is (presumably) based on the observation that low probability symbols $s$ are best allocated near the points $(n + \frac{1}{2})/f(s)$ for $n = 0, 1, \dots, Qf(s) - 1$, where $f(s)$ is the probability of the symbol $s$ and $Q$ is the table length.
We suspect that is it feasible to prove such a statement quantitatively in the limit as the length of the interval $\mathcal{I}$ of streamed asymmetric numeral systems, see Equation \eqref{eq:stream-shape-interval-gen}, goes to infinity.

Returning to the performance profiles in Figure \ref{fig:stream-Performance}, the observant eye may notice a slight performance drop of Dub\'e--Yokoo's algorithm \cite[\S III]{dube2019fast} for larger interval size. This is likely due to fact that we used (for all algorithms) the table generated for the small interval size extended periodically for the larger interval sizes. Hence, their table may be too specialised towards the dynamics of streamed asymmetric tables using a small table.

In conclusion, it is thus fair to say that Duda's second algorithm \cite[\S4]{DudaANS13}, Dub\'e--Yokoo's algorithm \cite[\S III]{dube2019fast}, and our more elaborate algorithms, shifted priorities and greedy discrepancy minimisation, see Algorithm \ref{alg:table-allocation-shifted-priorities} and \ref{alg:table-Greedy-discrepancy-minimisation}, respectively, perform all equally well in general.

\bibliography{refs}{} \bibliographystyle{alpha}
% -----------------
\end{document}